\documentclass[a4paper, 12pt]{article}
\usepackage[latin2]{inputenc}
\usepackage{graphicx}
\usepackage{color}
\usepackage{epsf}
\usepackage{amsmath}
\usepackage[mathscr]{eucal}
\usepackage{amssymb}
\usepackage{multicol}
\usepackage{subcaption}
\def \g  {\rm $\gamma$}

\begin{document}

\title{Paving the way to simultaneous multi-wavelength astronomy}
\author{M. J. Middleton\footnote{m.j.middleton@soton.ac.uk}, P. Casella, P. Gandhi, E. Bozzo, G. Anderson, N. Degenaar, \\I. Donnarumma, G. Israel, C. Knigge, A. Lohfink, S. Markoff, T. Marsh, N. Rea, \\S. Tingay, K. Wiersema, D. Altamirano,  D. Bhattacharya,  W. N. Brandt, S. Carey,\\ P. Charles, M. Diaz Trigo, C. Done, M. Kotze, S. Eikenberry,\\ R. Fender, P. Ferruit, F. Fuerst, J. Greiner, A. Ingram, L. Heil,\\ P. Jonker, S. Komossa, B. Leibundgut, T. Maccarone, J. Malzac, V. McBride,\\ J. Miller-Jones, M. Page, E. M. Rossi, D. M. Russell, T. Shahbaz, G. R. Sivakoff,\\ M. Tanaka, D. J. Thompson, M. Uemura, P. Uttley,  G. van Moorsel,\\ M. Van Doesburgh, B. Warner, B. Wilkes, J. Wilms, P. Woudt}
\date{}
\maketitle

\section{Introduction}

Whilst astronomy as a science is historically founded on observations at optical wavelengths, studying the Universe in other bands has yielded remarkable discoveries, from pulsars in the radio, signatures of the Big Bang at submm wavelengths, through to high energy emission from accreting, gravitationally-compact objects and the discovery of gamma-ray bursts. Unsurprisingly, the result of combining multiple wavebands leads to an enormous increase in diagnostic power, but powerful insights can be lost when the sources studied vary on timescales shorter than the temporal separation between observations in different bands. In July 2015, the workshop ``Paving the way to simultaneous multi-wavelength astronomy"  was held as a concerted effort to address this at the Lorentz Center, Leiden. It was attended by 50 astronomers from diverse fields as well as the directors and staff of observatories and spaced-based missions. This community white paper has been written with the goal of disseminating the findings of that workshop by providing a concise review of the field of multi-wavelength astronomy covering a wide range of important source classes, the problems associated with their study and the solutions we believe need to be implemented for the future of observational astronomy. We hope that this paper will both stimulate further discussion and raise overall awareness within the community of the issues faced in a developing, important field.

\section{Multi-wavelength astrophysics}

A great deal has been learnt from the combination of multiple observing bands and it is important for the sake of context, to review some of the insights which have been gained. In what follows we summarise (although we note that this is not exhaustive) some of the major science results obtained for a number of classes of astrophysical source and highlight important examples of what could be learnt from coordinated `simultaneous' observing.  

\subsection{Active galactic nuclei} The combination of emission from the accretion disc surrounding the supermassive black hole (SMBH, peaking in the optical/UV/soft X-rays depending on the black hole mass, accretion rate and spin) with that of the corona (emitting in the X-rays to soft \g-rays) and jets that cool via synchrotron (extending from the low frequency radio up to the near-IR) and inverse Compton (IC, emitting into the \g-rays) results in a spectral energy distribution (SED) that covers several decades in frequency. Studies of active galactic nuclei (AGN) that utilise the lever arm of multiple wavebands have provided insights into the nature of accretion onto their SMBHs including the coupling of inflow and outflow in the fundamental plane (where the energy in the jet is related to the SMBH mass and the accretion luminosity: Merloni, Heinz \& Di Matteo 2003; Falcke, K{\"o}rding \& Markoff 2004; Plotkin et al. 2012), the prevalence of line-driven winds and their location (revealed by blue-shifted absorption lines from the UV to X-rays, e.g. Kaastra et al. 2014; Tombesi et al. 2015; Parker et al. 2017), the size of the accretion disc (Fausnaugh et al. 2016) and, in the case of our nearest SMBH, Sgr A$^*$, how accretion operates at the very lowest (quiescent) rates. 

Besides the radiating plasma local to (or launched from near to) the SMBH, optical continuum emission overlaid with narrow and broad emission lines, their polarisation and the presence of a surrounding torus of dense molecular gas and dust -- which reprocesses incident X-rays into IR emission -- has led to the unified model of AGN (Antonucci \& Miller 1985; Antonucci 1993). Notably at large inclinations to the observer, the X-ray emission from the central source is diminished until high energies where the absorption opacity through the torus drops; high energy missions such as {\it INTEGRAL, Swift} and most recently {\it NuSTAR} have helped define the population of such Compton-thick AGN (typical estimates indicate a fraction of $\sim$ 20\% at redshifts $<$ 0.1: Koss et al. 2016) with the relative fraction providing a crude indicator of the covering fraction of the torus. 

The interaction between the optical broad-line emitting clouds and the central radiation source has allowed the mass of the SMBH to be measured via reverberation techniques (see the review of Peterson 2014). The latter has played a major role in helping understand the connection between the SMBH and the host galaxy, notably that the SMBH mass appears to be tightly correlated with the velocity dispersion of stars in the central bulge (the M$_{\rm BH}$-$\sigma$ relation, e.g. Gebhardt et al. 2000; Ferrarese \& Merritt 2000), the galaxy mass (Magorrian et al. 1998) and dark matter halo mass (e.g. Bogdan \& Goulding 2015). Such remarkable correlations demand that the SMBH and galaxy co-evolved through the process of `feedback' where the jets and winds revealed by multi-wavelength studies interact with the host, shutting off (or triggering) star-formation and dispensing momentum (see the review of Fabian 2012). The role and impact of feedback is beautifully illustrated in galaxy clusters where gas falls onto the central, largest SMBH with jets returning hot gas and thereby enriching the intra-cluster medium (ICM). The presence of shock fronts as well as synchrotron cavities inflated by jets indicate that large amounts of material and energy is being injected into the local environment (see Fabian 2012 and Figure 1 from McNamara et al. 2009).  

Figuring out how the relativistic jets in AGN are launched is a true multi-wavelength endeavour. The process can arguably be best studied in the so-called broad-line radio galaxies (BLRGs). Due to their inclination, this type of radio-loud AGN allows us to see both the emission from the AGN accretion flow and the jet emission at the {\it same} time. The evolution in the jet is tracked in the radio and \g-rays and the accretion disc in the UV/optical band, and finally the coronal emission can be observed in the X-ray band. Studying all these different emission components simultaneously in the BLRG 3C 120 has led to our current picture of jet formation in these sources (Marscher et al. 2002; Chatterjee et al. 2009; Lohfink et al. 2013) as shown in Figure 1 (Lohfink et al. 2013): a new jet knot is ejected as the inner part of the accretion disc is evacuated (perhaps magnetically disrupted or accreted on a rapid viscous timescale). This is then followed by the accretion disc resettling as the knot propagates further outward. Later, the process repeats itself steadily ejecting new knots into the jet every few months -- a timescale much faster than any expected viscous timescale of a thin accretion disc but easily observable. Verifying the existence of this jet cycle in other sources has been challenging because of the lack of simultaneous observations but some supporting evidence for similar behaviour has been found (Chatterjee et al. 2011). 

Whilst the feeding of the central SMBH in galaxy clusters may progress via mergers or accreted gas from the ICM, how AGN in field galaxies are generally fed is not entirely clear. However, once again, multi-wavelength observations can help: the misalignment of jet (revealed at radio to IR frequencies) and host galaxy's stellar disc (studied in the optical) would imply that mergers fuel the majority of sources (Kinney et al. 2000) whilst the misalignment of the sub-pc disc with the host galaxy as revealed via X-ray spectroscopy would seem to confirm that at least some proportion of the AGN population are not likely to be fed via a galactic disc (Middleton et al. 2016). 
\bigskip

\begin{figure}
    \centering
    \begin{subfigure}[t]{0.3\textwidth}
        \centering
        \includegraphics[width=70mm]{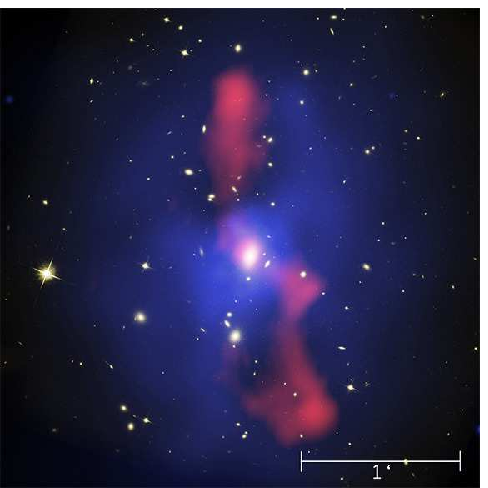} 
      
    \end{subfigure}
    \hfill
    \begin{subfigure}[t]{0.5\textwidth}
        \centering
        \includegraphics[width=80mm]{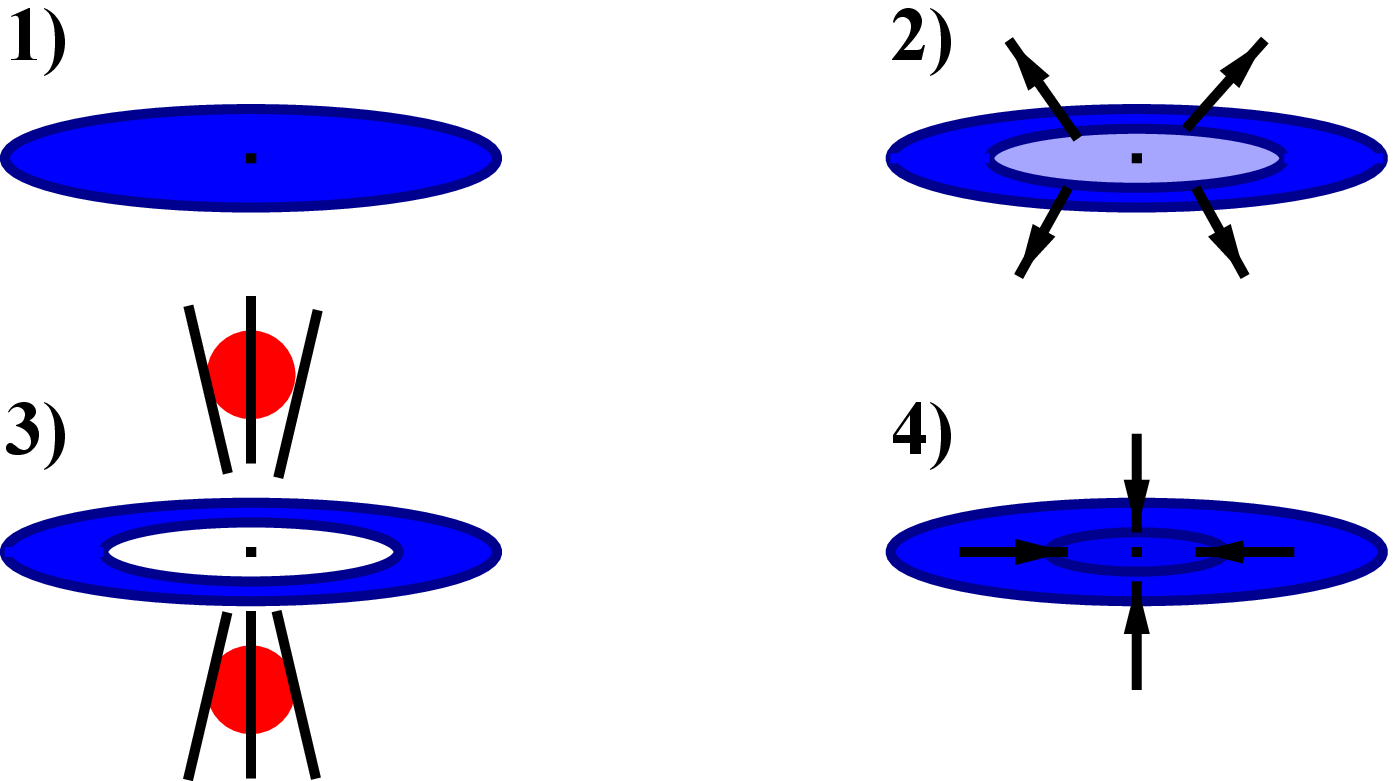} 
    
    \end{subfigure}
    \caption{Left: from McNamara et al. (2009) showing the inner 700 kpc of the cluster MS0735.6+7421 with the hot gas in X-rays (blue), central galaxy in I-band (white), and jets in the radio (red). Such jets redistribute energy released via accretion onto the central SMBH and are responsible for the heating of the gas (see the review of Fabian 2012). Right: from Lohfink et al. (2013) showing the proposed jet cycle in AGN (based on 3C~120) where at 1) the accretion disc is full, 2) the inner disc becomes unstable, 3) a jet is formed, 4) the disc refills. (see Lohfink et al. 2013 for more detail). Probing this cycle in depth and in other sources is expected to require simultaneity on the order of weeks-months (faster than might be predicted for the viscous timescale in a thin disc).}
\end{figure}

\noindent {\bf What coordinated observing could reveal:}
\smallskip 

Given the discovery of an apparent disc-jet coupling cycle occurring far faster than expected from viscous timescales in 3C~120 (Lohfink et al. 2013), simultaneous monitoring on the timescales of weeks-months of such systems across IR-optical-UV-X-rays holds the promise of revealing how jets from such AGN are launched. 

Whilst we cannot realistically hope to resolve the sub-pc accretion disc in AGN with current instrumentation, the inflow has recently been mapped using time lags which probe the reprocessing of variable illumination spanning the optical to X-rays (e.g. McHardy et al. 2014; Edelson et al. 2015). This is a powerful technique and requires simultaneity between bands on timescales of less than that of the time lag; as this is a light travel time, simultaneity on timescales of $<$ hours is often required (extending to day timescales for the largest size scales). 

\begin{figure}
\begin{center}
\begin{tabular}{l}
 \epsfxsize=14cm \epsfbox{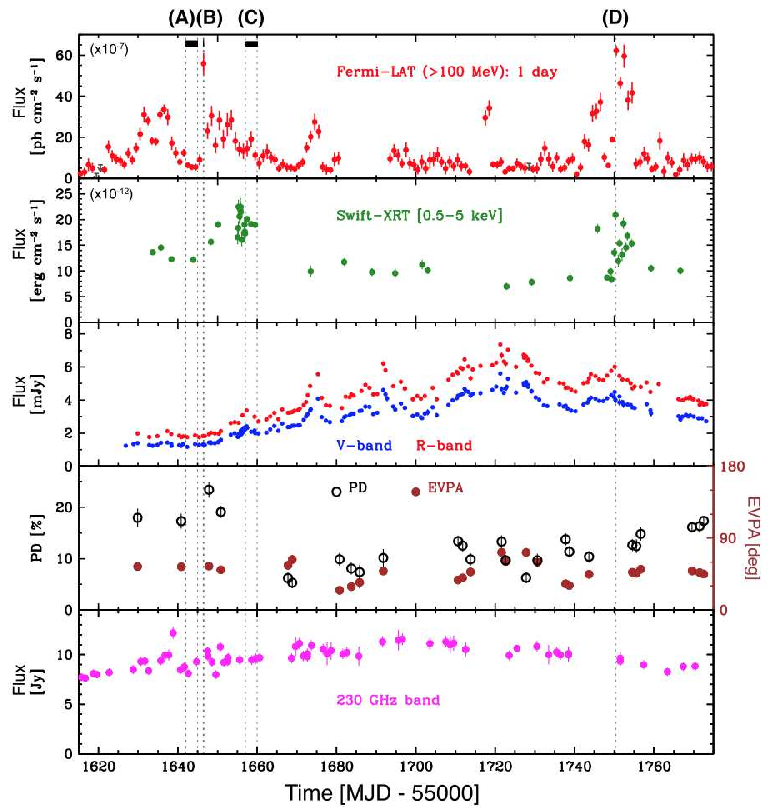}
\end{tabular}
\end{center}
\vspace{-0.2cm}
\caption{Multi-wavelength light curves of the blazar, 3C 279, in the high activity state detected between 2013 December and 2014 April. From the top to bottom: \g-ray data above 100 MeV from {\it Fermi}-LAT; X-ray (0.5-5 keV) data from {\it Swift}-XRT; optical data from {\it Swift}-UVOT, SMARTS and Kanata; optical polarization measurements from Kanata (degree and electric vector polarization angle) and mm (230 GHz) flux densities from SMA and ALMA. This dataset points out the importance of the long-term monitoring in different energy bands to investigate the presence (or lack) of correlated variability between the synchrotron and IC peaks. In addition, ToOs in the most intense activity periods are crucial to detect variability on shorter timescales. Adapted from Hayashida et al. (2015).}
\label{fig:l}
\end{figure}

\subsubsection{Blazars}

In the framework of the AGN unification scheme, blazars are radio--loud systems with jets oriented close to the observer's line-of-sight. Blazers demonstrate emission extending from radio to \g-rays (even up to TeV energies) with a double-humped SED whose origin is believed to be the boosted non-thermal emission from the relativistic jet. The synchrotron emission from energetic electrons in a tangled magnetic field accounts for the low energy bump of the SED. The high energy emission is commonly explained in leptonic models due to IC scattering off relativistic electrons in the jet by a seed photon field which can originate from the synchrotron emission itself (synchrotron self-Compton, SSC) or from a source external to the jet such as the disc, the broad-line region or the molecular torus. 

Noticeable advances in the data collection for several blazars have been achieved during the last nine years, mainly due to unprecedented observations in \g-rays by {\it AGILE} and {\it Fermi}. The continuous monitoring of the \g-ray sky performed by {\it Fermi} with its large area telescope (LAT)  revealed that thousands of blazars emit at very high energies. This increased sample has stimulated the synergy among different groups working with ground-based telescopes and satellites from radio up to TeV energies e.g. the GASP-WEBT network (created to support both {\it AGILE} and {\it Fermi} observations of blazars with radio/IR/optical follow-up) and the SMARTs blazar programme. 

High-resolution radio images at 43 GHz and optical polarization monitoring available for a sample of blazars detected in \g-rays are providing important diagnostics for the mechanisms of the injection of material into the jet and the topology of the magnetic field lines. In addition, \textit{Swift} is playing a key role in the X-rays thanks to the rapid response to `target of opportunity' requests (ToOs) and the advantage of immediately public data. These data provide an important bridge to connect source properties at lower energy with those in \g-rays.

The multi-wavelength coordinated blazar programs described above have challenged the standard model; it was usually thought that the bulk of the \g-ray emission was produced close (sub-parsec scale) to the SMBH, but some multi-wavelength data suggest that it may instead be produced at larger distances (pc scale from the SMBH, Marscher et al. 2011). Indeed, all of the presently available diagnostics lead to a scenario which is very complex and difficult to interpret, mainly because of the difficulties in reconciling the following: 
\begin{itemize}

\item A lack of a sharp cutoff in the \g-ray spectrum above 20 GeV/(1+z) (Pacciani et al. 2014), expected as a result of \g-ray absorption by photon-photon pair production, in turn due to the Lyman alpha line and recombination continuum produced in the broad line region (Poutanen \& Stern 2010).
\item The radio-\g-ray connection (Jorstad et al. 2013) and polarization swings (Abdo et al. 2010) suggest that the low and high-energy emitting regions are located at larger distances, i.e. pc from the SMBH, in coincidence with the bright compact structure at 43 GHz (the `core') observed by VLBA at millimeter wavelengths to be at the upstream end of blazar jets.
\item The short timescale variability ($\sim$ minutes) observed in powerful \g-ray flares (e.g. 3C~279, Ackermann et al. 2016) suggest very compact emitting regions at large distances from the SMBH. In order to avoid having too compact a region (with the problem of \g-ray absorption due to $\gamma-\gamma$ pair production), the bulk Lorentz factor of the emitting region must have a value close to or even larger than $\Gamma = 50$ (Begelman, Fabian \& Rees 2008). These values of $\Gamma$ appear larger than any value inferred from measurements of superluminal motion in AGN. Therefore this extreme variability could indicate that flares occur in subregions of the jet, thus requiring physical mechanisms alternative to diffusive shocks, efficient on very small spatial lengths (e.g. magnetic reconnection, turbulent cells).
 
\end{itemize}

\bigskip

\noindent {\bf What coordinated observing could reveal:}
\smallskip 

Where the high energy emission is produced and the responsible physical process in blazars is still unknown mainly because of the difficulty in obtaining simultaneous data at different epochs and/or
different states for a large number of sources. The situation is even more complicated when we consider that
the {\it same} source can display {\it different} behaviours in their \g-ray activity (e.g. 3C 454.3). In this respect,
having long term monitoring of a sample of blazars in radio, optical and \g-rays, with at least
a week sampling and simultaneity on day timescales, could add important information about the duty cycle of the blob injection in the
jet and the mechanism responsible for the flaring emission (see Figure 2).

\subsubsection{Tidal disruption events}

\begin{figure}
\begin{center}
\begin{tabular}{l}
 \epsfxsize=14cm \epsfbox{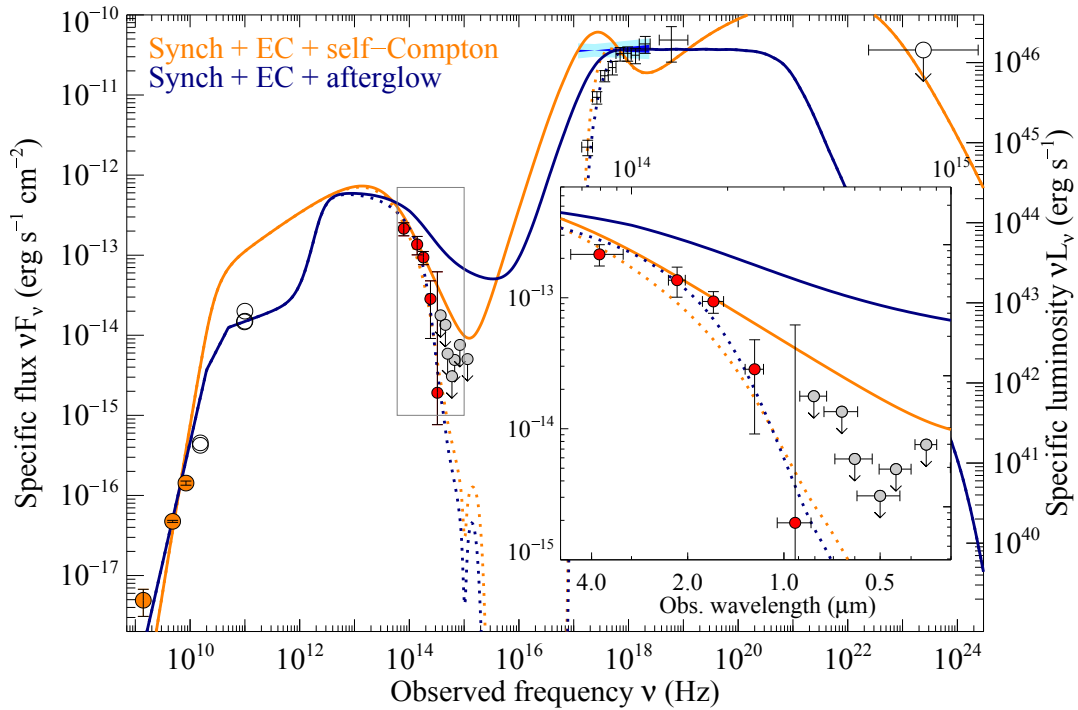}
\end{tabular}
\end{center}
\vspace{-0.2cm}
\caption{From Bloom et al. (2011). SED of the jetted TDE, Swift J1644+57 2.9 days after detection and created from the combination of measurements and data from published circulars. Clearly, any attempt to understand the energetics of such systems requires an extremely broad-band view.}
\label{fig:l}
\end{figure}

Whilst AGN in both field galaxies and in the centres of clusters appear at present to be fed over relatively long (typically $>$ human) timescales, the discovery of short-lived, high amplitude flaring in otherwise quiescent galaxies indicates the presence of tidal disruption events (TDEs) where a star is torn apart by the strong tidal forces of the central SMBH (see Rees 1990 and the review of Alexander 2012).

TDEs are traditionally divided into two subclasses of `jetted' and `non-jetted' (see the review of Komossa et al. 2015). The latter are characterised by high-amplitude flares in the X-ray, optical and UV. Those discovered in the X-rays (many by the {\it ROSAT} all-sky survey, e.g. Bade et al. 1996; Komossa \& Greiner 1999), are characterised by soft X-ray luminosities in excess of 10$^{44}$ erg/s which fade according to the predicted t$^{-5/3}$ law (Rees 1988; Evans \& Kochanek 1989). TDEs seen in the optical (by SDSS, Pan-STARRS and Palomar Transient Factory -- PTF) and UV bands (by {\it GALEX}) have considerably softer emission (with flare black-body temperatures of 10$^{4}$~K rather than 10$^{5}$~K seen in X-ray TDEs) and can be distinguished from supernovae by their characteristic lightcurves and SEDs.

The first (and to-date, best) example of a `jetted' TDE is that of Swift J164449.3+573451 (see Bloom et al. 2011, Figure 3). Discovered as a result of the source triggering the {\it Swift} Burst Alert Telescope (BAT), the X-ray emission reached a remarkable 10$^{48}$ erg/s equivalent to $\ge$1000$\times$L$_{\rm Edd}$ (where $L_{\rm Edd}$ is the Eddington luminosity) for SMBH mass estimates of $<$10$^{7}$ M$_{\odot}$ (Burrows et al. 2011; Bloom et al. 2011). Supported by the presence of bright, variable radio emission (e.g. Zauderer et al. 2011) associated with synchrotron cooling, the favoured explanation is that a TDE and resultant -- likely super-Eddington -- accretion flow triggered the production of a jet (with geometric beaming of the X-ray emission by the wind  - see King 2009 and Kara et al. 2016) - probably leading to an overestimate of the accretion rate. 

It is worth noting that a recent result indicates that the adopted definition of TDEs as `jetted' or `non-jetted' is probably misleading; radio jets have now been
detected from a `non-jetted' event and are likely common to all TDEs at early times (van Velzen et al. 2016). 

Predicted rates for TDEs are in the range 10$^{-4}$-10$^{-5}$ yr$^{-1}$galaxy$^{-1}$ (e.g. Esquej et al. 2008; Luo et al. 2008; Gezari et al. 2008; Maksym et al. 2010; Khabibullin \& Sazanov 2014; van Velzen \& Farrar 2014) although such rates can be boosted by galaxy mergers (e.g. Chen et al. 2009). With the advent of new instruments such as the Large Synoptic Survey Telescope (LSST) and their precursors, we can expect a generational leap in the number of detections, with multi-wavelength studies shining a light onto the processes behind these powerful events (and which may allow such important developments as reverberation mapping of the central regions of galaxies and understanding the launching mechanism for jets in this extreme accretion regime).
\bigskip

\noindent {\bf What coordinated observing could reveal:}
\smallskip 

In order to study TDE jet formation (and how this compares between the thermal and non- thermal subclasses) rapid multi-wavelength follow-up (within a few days) mainly in radio and X-rays is required to study the initial phase of the event and the onset of the jet. In order to study substructures in the lightcurve, simultaneity on sub-day timescales is expected to be required. Finally, it is important to stress the importance of long-term monitoring of TDEs in order to identify the point at which the jet switches off and how this connects with the properties of the accretion flow (see Donnarumma \& Rossi 2015).

\subsubsection{Sgr A$^{*}$}

\begin{figure}
\begin{center}
\begin{tabular}{l}
 \epsfxsize=14cm \epsfbox{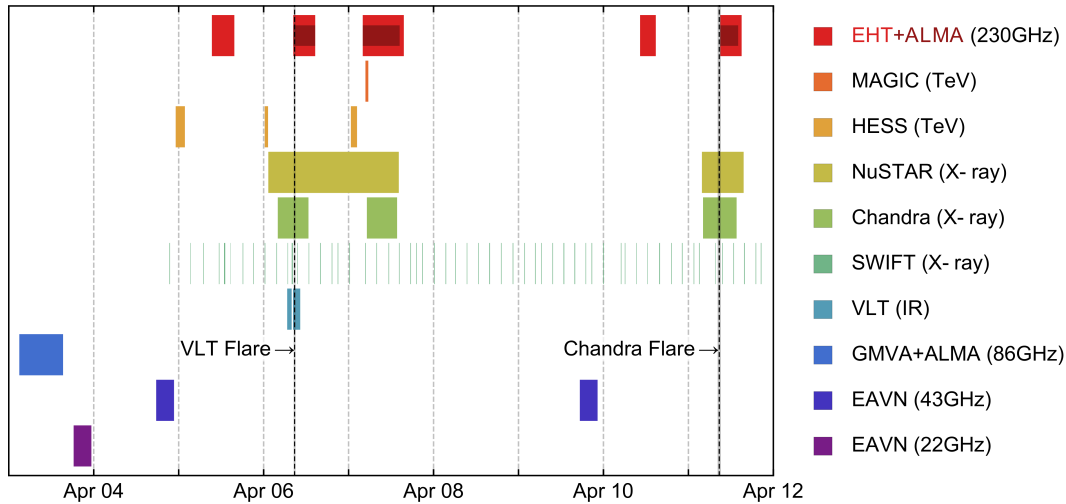}
\end{tabular}
\end{center}
\vspace{-0.2cm}
\caption{An example of the observing program coordinated around the first Event Horizon Telescope (EHT) observations of Sgr A*, across radio, sub-mm, IR, X-ray to VHE \g-rays. Such coverage is vital to probe the plasma conditions during the mm-VLBI imaging and constrain quantities such as optical depth, in order to help with the modelling and interpretation, and serves as an example of the levels of successful coordination that can be achieved by existing means. It is worth noting that even in this well-coordinated campaign, there were substantial challenges: notably the lack of advanced warning of observations required scheduling of ToOs to ensure optimal (if not overlapping) coverage. Figure credit: Michael Johnson, on behalf of the EHT Multiwavelength WG, used with permission of EHTC.
}
\label{fig:l}
\end{figure}

Our nearest SMBH, Sgr A* accretes at very low rates referred to as `quiescence' (around 10$^{-9}$ of its Eddington limit). Due to its proximity, studying this source can naturally provide us with the best view of accretion at such rates (and may well be analogous to quiescence in BHXRBs, see Connors et al. 2017, and the following section). There have been many multi-wavelength studies of Sgr A* which shows both persistent emission and flares (that bring its emission properties closer to those of other low luminosity AGN and quiescent black hole X-ray binaries (BHXRBs), e.g. Markoff et al. 2015). The former is characterised by flat/inverted, optically-thick synchrotron emission typical of compact radio cores in AGN, rising more steeply towards the mm/submm, with typically $\sim$20\% fractional flux variability in the cm/mm bands (from the VLA, IRAM, Nobeyama and BIMA telescopes; see, e.g., Falcke et al. 1998), typical of other self-absorbed, low-luminosity AGN cores (Ho et al. 1999), while the quiescent thermal X-ray flux has been stable since its discovery almost 20 years ago (see e.g. Genzel et al. 2010, Yuan \& Narayan 2014). This thermal component originates near the Bondi radius and has been resolved on scales of $\sim10^{5}$ $r_g$ (where $r_g = GM/c^{2}$ and $M$ is the mass of the black hole in SI units; Wang et al. 2013).  

At frequencies above the so-called `submm bump' (Falcke et al. 1998), the spectrum becomes optically-thin and turns over, revealing a continuous range of variability in the two available windows of the IR and X-ray bands. X-ray observations show roughly daily non-thermal flares with amplitudes from a few to hundreds of times the quiescent value. While X-ray flares are {\it always} associated with flares in the IR, some IR flares are {\it not} associated with flares in the X-rays. This can be explained by the quiescent non-thermal emission of Sgr A* falling below the quiescent thermal flux from larger scales, which effectively blankets any smaller X-ray flares. Generally the IR/X-ray flares are simultaneous, although new campaigns with {\it Spitzer} (Hora et al. in prep.) call this into question. There are only two viable mechanisms to explain the flares: direct synchrotron emission from the same population of particles creating the submm bump, or IC (either SSC or from another photon field). Despite several multi-wavelength campaigns, it has been a persistent challenge to obtain the kind of high-quality, simultaneous submm/IR/X-ray observations during a flare that would determine once-and-for-all which mechanism is responsible. For instance, based on non-simultaneous flares, Dibi et al. (2014) showed that the most physical model would be synchrotron emission. Conversely, recent studies of the statistical properties of the flares in the IR/X-rays (Witzel et al. 2012, Dodds-Eden et al. 2011, Neilsen et al. 2015) show two different distribution functions, which would argue against such a simple scenario (Dibi et al. 2016). Interestingly, the first full Event Horizon Telescope observations in 2017 were coordinated with both IR and X-ray observations (see Figure 4), and one moderate X-ray flare was detected simultaneously with both the {\it Chandra X-ray observatory} and {\it NuSTAR}, which may help to finally pinpoint the flare origin and emission mechanisms.
\bigskip

\noindent {\bf What coordinated observing could reveal:}
\smallskip 
As stated above, it has been a persistent challenge to obtain high-quality, simultaneous sub-mm/IR/X-ray observations during a flare to unambiguously determine their origin. The required simultaneity is $\sim$hour, however, unlike other sources where there are populations available to study, Sgr A* is unique and presents unique challenges. Notably, various instruments located around the world have to wait for the Earth's rotation to bring Sgr A$^{*}$ into view which may take longer than a typical flare duration (e.g. the source is not visible by Keck and VLA in the same visibility window making simultaneity in those bands challenging to obtain).

\subsection{Compact object binaries} 

Given the hundreds of billions of stars within the Milky Way, we should expect hundreds of {\it millions} of compact objects resulting from gravitational collapse at the end of a star's lifetime: stellar mass black holes (BHs), neutron stars (NSs) and white dwarfs (WDs). Whilst many of these may be isolated (van den Heuvel 1992), bright X-ray emission with characteristic temperatures consistent with an accretion disc around compact objects with masses typically $<$12 M$_{\odot}$ (Reid et al. 2014) provides overwhelming evidence for the presence of actively accreting systems.

\begin{figure}
\begin{center}
\begin{tabular}{l}
 \epsfxsize=15cm \epsfbox{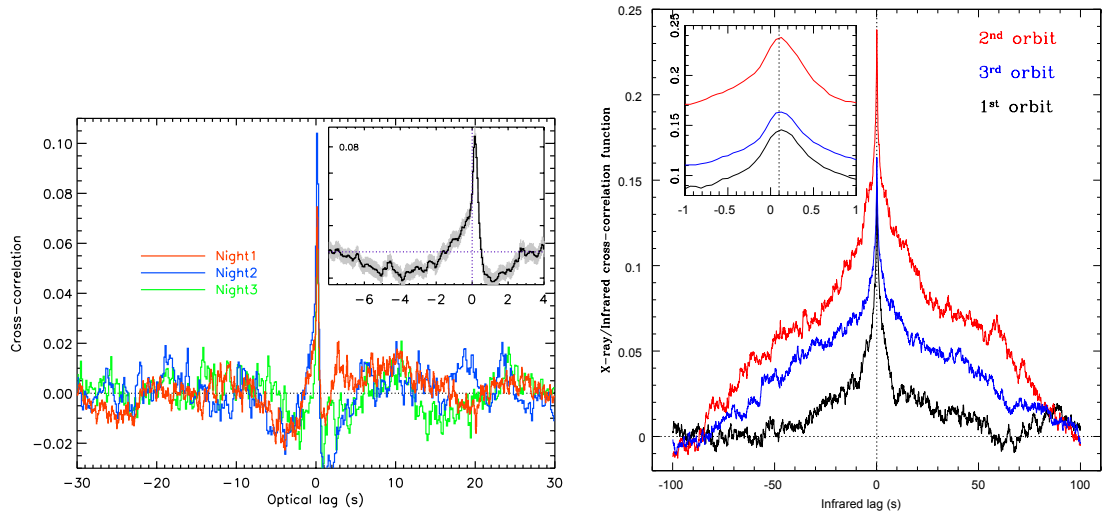}
\end{tabular}
\end{center}
\vspace{-2.5cm}

\caption{Cross-correlation functions of GX 339-4 constructed from simultaneous observations in X-rays ({\it RXTE}) and either optical (r' band, left, taken from Gandhi et al. 2008) or IR (Ks band, right, taken from Casella et al. 2010). A positive delay (in these cases peaking at $\sim$100-150 ms) implies that the optical/IR emission lags that in the X-rays (the insets show a zoom-in in order to clearly illustrate the delay). Whilst the two campaigns were not simultaneous, the IR/X-ray campaign on this (BH)LMXB identified sub-second variability from the jet, while the optical/X-ray campaign identified the presence of multiple emitting components varying on different timescales. Such campaigns and analysis are vital for determining the interplay and geometry of the accretion flow in such systems (see also Uttley \& Casella 2014)}
\label{fig:l}
\end{figure}

\subsubsection{Black hole binaries}

The majority of transient black hole X-ray binaries (BHXRBs) have low mass donor stars (the systems can be referred to as low-mass X-ay binaries: (BH)LMXBs) and enter occasional outbursts lasting weeks-months, likely triggered by a thermal-viscous disc instability (for a recent review see Lasota 2015 and the review of Done, Gierli{\'n}ski \& Kubota 2007 for observational details). Most (but not all, e.g. Smith et al. 2001) sources appear to follow a similar track in X-ray spectral hardness (the ratio of fluxes between hard and soft bands) and accretion rate/brightness, referred to as a hardness-intensity diagram (HID, e.g. Maccarone \& Coppi 2003; Fender, Belloni \& Gallo 2004; Belloni \& Motta 2016). It is thought that similar spectral states are also present in AGN (see K{\"o}rding Jester \& Fender 2006; Middleton et al. 2007; Jin et al. 2011) although the evolutionary timescales are expected to be far longer (typically decades or more) making studies of analogous evolution impractical.

Upon outburst, a given (BH)LMXB will rise up the `hard branch' of the HID where the X-ray spectrum is dominated by hard power-law emission up to very high (hundreds of keV)  energies due to Comptonisation by hot, free electrons (possibly a combination of thermal and non-thermal populations) reaching Eddington ratios of up to $\sim$70\%. In this radiatively inefficient state, X-ray emission from the accretion disc is weak (see Wilkinson \& Uttley 2009) but due to reprocessing of coronal photons in the outer disc, the optical emission can be substantial (see e.g. van Paradijs 1981; van Paradijs \& McClintock 1994; Gierli{\'n}ski, Done \& Page 2009; Curran, Chaty \& Zurita Heras 2012). Steady, compact jets (probably with bulk Lorentz factors, $\Gamma$ $<$ 2: Fender, Homan \& Belloni 2009) are known to accompany this accretion state and are identified via synchrotron cooling of relativistic electrons and emit across several orders of magnitude in frequency (e.g. Russell et al. 2014). The jet spectrum is thought to evolve, with the break from optically thick (synchrotron self-absorbed) to optically thin moving as a function of time and spectral state as the electron population cools (e.g. Russell et al. 2014) whilst strictly simultaneous observations have revealed that the optically-thick to thin break can be variable on timescales of minutes to hours (Gandhi et al. 2011) and does not follow simple jet scalings (Russell et al. 2013). From the brightest hard states, a source will transition to an intermediate state (hard then soft) where the relative fraction of direct X-ray emission from the disc increases and the power-law emission softens. At some point, the jet transitions to high bulk Lorentz factors (typically quoted as $\Gamma\sim$2-5, see Vadawale et al. 2003; Fender, Belloni \& Gallo 2004 but also Fender et al. 2003; Miller-Jones et al. 2006) and can be spatially resolved as synchrotron emitting ejecta (e.g. Mirabel \& Rodriguez 1994) which are brighter than the hard state jets and evolve according to adiabatic expansion. When such jets are produced, the source is termed a `microquasar' (it is thought that most BHXRBs probably enter such a state). Eventually the source enters the softest state where the disc emission dominates, and the power-law becomes flat (in log $\nu F\nu$ vs log $\nu$ space) and non-thermal (extending beyond several hundred keV: Gierli{\' n}ski et al. 1999; McConnell et al. 2000). In this state, powerful winds are found to be present via X-ray absorption lines (Miller et al. 2006; Neilsen \& Lee 2009; Ponti et al. 2012) and it has been suggested that such mass loss may hinder the production of jets which are observed to be absent (although see Wu et al. 2001; Homan et al. 2016). Typically over the course of months (commensurate with the viscous timescale of the outermost radius of the disc), the source decays in the soft state down to a few percent of the Eddington limit (derived from the X-ray luminosity, see Maccarone 2003) before returning into faint counterparts of the intermediate and hard states.

Whilst the presence of a radiative inflow (be it inefficient or efficient) via a disc and the launching of jets/winds is well established, the interplay between these different physical aspects of accretion is still not entirely clear. However, in the X-ray hard state, the tight correlation between radiative jet power, X-ray luminosity and black hole mass - and which extends from BHXRBs through to AGN (the fundamental plane) - indicates that the inflow and (in this case collimated) outflow are causally connected (see Merloni, Heinz \& Di Matteo 2003; Falcke, K{\"o}rding \& Markoff 2004; Plotkin et al. 2012). The correlation between the variability of the inflow (via observations in the X-ray band) and of the jet (seen in radio-optical - see Figure 5) confirm a causal connection (see Kanbach et al. 2001; Gandhi et al. 2008, 2010; Casella et al. 2010) with anti-correlated optical/X-ray behaviour observed on timescales of $\sim$1-10 s interpreted as arising from SSC in a central hot flow (Kanbach et al. 2001, Gandhi et al. 2010, Durant et al. 2008, 2011, Veledina et al. 2011, Pahari et al. 2017). In addition to `broad-band' variability, more coherent timing  features in the form of quasi-periodic oscillations (QPOs) have been detected at low frequencies ($<$ 1~Hz) in the optical and IR bands during outbursts (often simultaneous with X-ray QPOs); whilst relatively under-studied, these QPOs can provide important clues to the dynamics and structure of the hot flow (Hynes et al. 2003, Gandhi 2009, Kalamkar et al. 2016, Ingram et al. 2016, 2017; Veledina et al. 2015). Although the sample size remains small (see the section on ULXs) it has been proposed that the more powerful, discrete ejections (also called ballistic jets), launched during the transition from the hard to the soft state and detected in the radio band, are powered by `tapping' the black hole spin (via the Blandford-Znajek effect: Blindfold \& Znajek 1977; see Steiner et al. 2013 and McClintock et al. 2014 for the observational results), the latter measured by X-ray techniques (see Middleton 2016 for a review). Whilst this is still debated (see Russell et al. 2013), multi-wavelength observations are of great importance for determining (either directly or indirectly) how such jets are launched.

Between outbursts, the mass accretion rate is typically orders of magnitude lower (such that L $\le$ 10$^{-6} \times$ L$_{Edd}$) and, as with the accretion flow onto Sgr A$^{*}$, is termed `quiescent'. At such rates, the accretion flow is thought to be cooled via advection or outflows (or a combination). In practice, the very faint X-ray emission detected in such states requires that at least some of the energy intrinsic to the accretion flow is released, but the manner in which this occurs is unknown with suggestions ranging from radiatively inefficient jets (Fender, Gallo \& Jonker 2003) to strong outflowing winds (Blandford \& Begelman 1999). The problem is compounded, as the vast majority of quiescent BHXRBs emit at levels impractical for study, and so there is little observational evidence to distinguish between competing models. V404 Cygni is one notable exception having a flux $>$ 1000 times higher at X-ray and radio energies (even though typically at 1$\times$10$^{-6}$ L$_{Edd}$) than the average of the population due to its proximity (2.4 kpc: Miller-Jones et al. 2009) and relatively long orbital period (155.3hrs, e.g. Casares \& Jonker 2014). Full multi-wavelength coverage of this source is possible, allowing the properties of the emission to be well-studied and its origin determined. The most complete multi-wavelength analysis of V404 Cygni in quiescence (Hynes et al. 2009) has revealed that the X-ray (0.3-10 keV) band is entirely dominated by a power-law component with a photon-index $>$ 2 (consistent with all other analyses) whilst at radio frequencies, the emission is a flat power-law (i.e. the flux, $S_{\nu} \propto \nu^{alpha}$ where $\alpha$ = 0 in this case). Although the position of the break from synchrotron self-absorbed to optically thin has not been constrained in quiescence due to contamination by the sub-giant companion star (Gallo et al. 2007, although it has been constrained in the decay of the hard state: Russell et al. 2013), this may suggest that the emission originates in synchrotron cooling either in an outflow generated in the advection-dominated accretion flow (ADAF: Ichimaru 1977; Narayan \& Yi 1994; 1995) or in a (radiatively inefficient) jet (e.g. Fender, Gallo \& Jonker 2003). Multi-wavelength studies of this source in outburst are discussed in detail in Section 4. Another source that can be well studied in quiescence is the Galactic (BH)LMXB A0620-00 (e.g. Gallo et al. 2007), which is the lowest luminosity black hole studied other than Sgr A* (see preceding sub-section). Even closer than V404 Cygni ($\sim1$ kpc) it seems to behave in very similar manner to Sgr A*, and the same mass-scaled model can fit both sources (Connors et al. 2017), providing clues to the emission mechanisms and indicating a weakening of particle acceleration efficiency at low accretion rates.  The presence of thermal synchrotron originating near the black hole in quiescence (see also Shahbaz et al. 2013), similar to Sgr A$^{*}$'s submm bump, may be the signature of the jet base or magnetised corona, revealed by the lack of an efficiently accelerated population of electrons (i.e. a power law of emission).
\bigskip

\noindent {\bf What coordinated observing could reveal:}
\smallskip 

The now robust evidence for variability in the accretion flow (studied in the X-rays) being transferred into and along the outflows (observed at longer wavelengths: Eikenberry et al. 1998; Casella et al. 2010; Gandhi et al. 2010; Lasso-Cabrera \& Eikenberry 2013) will make simultaneous, high time resolution, multi-wavelength monitoring of BHXRBs a powerful tool to study disc-jet coupling, providing strong constraints on the jet launching mechanisms. The timescales we need to investigate are typically those of viscous propagation into the jet whilst the evolution of the jet spectrum will be determined by the cooling timescale (which may be that for thermally driven adiabatic expansion and so depend on the position from the jet launching/acceleration zone). The fastest viscous timescales in BHXRBs (those at the ISCO) can be extremely short (of the order of seconds or less), whilst time/phase lags between physical components can be even shorter (milliseconds or less). To study the geometry via the transfer function (see the review of Uttley et al. 2014) is therefore extremely challenging in these objects (although not necessarily in AGN where timescales are longer - see Reynolds et al. 1999) however, determining the energetic {\it coupling} between physical components is already possible should effective simultaneity be achieved.

\subsubsection{Neutron star binaries}

As with BHXRBs, the radio emission from accreting neutron star binaries (NSXRBs) traces outflows in the form of a jet and is correlated with the X-ray emission from the accretion flow (Migliari \& Fender 2006). However, there are some indications that the radio/X-ray correlation has a steeper index for neutron stars than for black holes (Migliari et al. 2003, 2006; Tetarenko et al. 2016), which has been interpreted as the difference between radiatively efficient (neutron star) versus non-efficient (black hole) accretion (Fender \& Kuulkers 2001; Migliari et al. 2006; Deller et al. 2015). Whereas in BHXRBs, the radio emission from a steady jet is quenched in the soft X-ray spectral state, a few neutron star (NS)LMXBs are still detected in the radio band during their soft states (Migliari et al. 2004), although in some cases the radio emission is strongly suppressed compared to hard X-ray spectral states (Miller-Jones et al. 2010; Migliari et al. 2011). 

Quasi-simultaneous X-ray and optical/IR observations of (NS)LMXBs show that the luminosity in the different wavebands is correlated over many orders of magnitude in X-ray luminosity (Russell et al. 2006; Maitra  \& Bailyn 2008; Zhu et al. 2012, see also Figure 6). The functional shape of these correlations suggest that in most (NS)LMXBs the optical/IR emission during hard X-ray spectral states is dominated by X-ray reprocessing (e.g. van Paradijs 1981; van Paradijs \& McClintock 1994), with possible contributions from the jets (at high luminosity) and the viscously-heated disc (Russell et al. 2006). Different types of X-ray outbursts appear to result in different correlations between the optical/IR and X-ray emission, suggesting fundamental differences in the accretion flow morphology (Maitra et al. 2008). In some (NS)LMXBs, the correlation between the optical/IR and X-ray bands is more similar to black hole systems (see Gandhi et al. 2008, 2010; Casella et al. 2010) and has been interpreted as evidence of radiatively-inefficient accretion in some (NS)LMXBs, e.g. due to the magnetic field propelling material away (Patruno et al. 2016).

A small number of (NS)LMXBs have been found to display prominent \g-ray emission. This mostly concerns members of a sub-class of neutron star that are able to transition into a radio pulsar during non-accreting states (Hill et al. 2011; Stappers et al. 2014; de Ona Wilhelmi et al. 2016). When these objects shine brightly in \g-rays (they are simultaneously radio bright due to a compact jet), they are analogous to a (NS)LMXB but with a much lower X-ray luminosity than is commonly seen for active systems. This has led to the suggestion that the \g-rays are generated by shocks formed when the accretion stream runs into the neutron star magnetosphere (Papitto, Torres \& Li 2014; Papitto \& Torres 2015) that likely truncates the accretion disc (Patruno et al. 2014). 
\bigskip

\begin{figure}
\begin{center}
\begin{tabular}{l}
 \epsfxsize=14cm \epsfbox{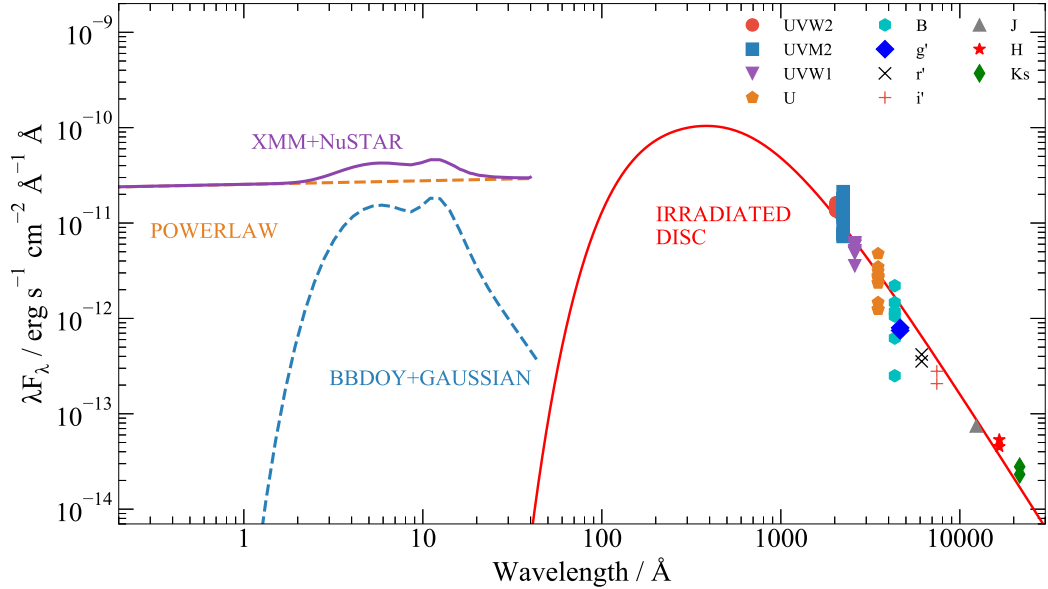}
\end{tabular}
\end{center}
\vspace{-0.2cm}
\caption{SED of the Galactic (NS)LMXB IGR J17062-6143, constructed from non-simultaneous Magellan near-IR (J, H, K), Faulkes optical (g', r', I'), {\it Swift} optical/UV (b, u, uvw1, uvw2, um2), {\it XMM-Newton} soft X-ray and {\it NuSTAR} hard X-ray (purple curve) observations. The {\it Swift} UV/optical observations were obtained at different epochs and illustrate the level of variability that is present on day-month timescales. Such variability causes difficultly in accurate SED modelling and emphasises the need for simultaneous multi-wavelength observations. The SED model shown in this plot consists of several components: black body emission (from the NS surface) and disc reflection in soft X-rays  (blue dashed curve), a non-thermal power-law at hard X-rays (orange dashed curve), and an irradiated disc at UV-optical-near-IR wavelengths (red solid curve). This plot was adapted from Hern{\'a}ndez Santisteban et al (submitted). Credit: Juan Hern{\'a}ndez Santisteban.}
\label{fig:l}
\end{figure}

\noindent {\bf What coordinated observing could reveal:}
\smallskip 

Performing simultaneous multi-wavelength observations of (NS)LMXBs would be a very powerful tool to further our understanding of the dynamics of the accretion disc and associated outflows, the possible interaction between the magnetic field of the NS and the disc, and how the accretion morphology is changing as the accretion rate drops.  For instance, cross-correlating X-ray, UV, optical and near-IR light curves would provide the opportunity to search for correlations and delays between different wavelengths, which is a powerful tool to understand the origin of the different emission components (Russell et al. 2006; Maitra et al. 2008; Rykoff et al. 2010; Bernardini et al. 2013, 2016; Patruno et al. 2016). The typical timescales involved range from seconds to weeks. Furthermore, (NS)LMXBs may show repeating type-I X-ray bursts, resulting from unstable nuclear burning of the material accreted onto the NS surface (for reviews see Lewin et al. 1995; Strohmayer \& Bildsten 2006). Such bright X-ray flashes have a typical duration of seconds--minutes (although some can last as long as hours), can repeat up to every few hours, and lead to optical/UV bursts that are likely due to reprocessing of the X-ray emission in the accretion flow (Hackwell et al. 1979; Matsuoka et al. 1984; Hynes et al. 2006). Simultaneous (on sub-second timescales) measurements of the direct type-I X-ray burst emission and its reprocessed optical/UV signature will be a powerful tool to understand the accretion geometry and disc structure, whilst studying changes associated with the {\it disruption} of the accretion flow due to type I bursts (requiring simultaneity on the timescales of the bursts) will also provide valuable insights (see e.g. Maccarone \& Coppi 2003; Keek, Wolf \& Ballantyne 2016).

\begin{figure}
\begin{center}
\begin{tabular}{l}
        \includegraphics[width=100mm]{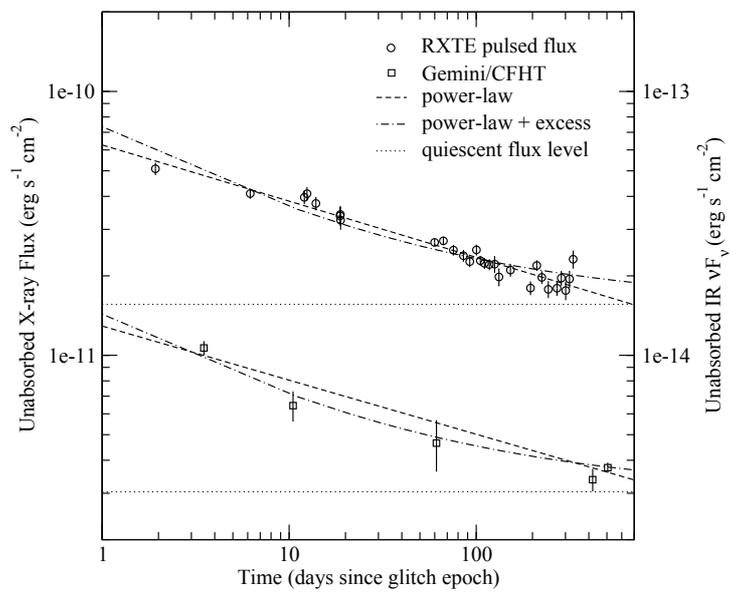} 
\end{tabular}
\end{center}
\vspace{-0.2cm}
    \caption{From Tam et al. (2004): unabsorbed X-ray flux and near-IR (Gemimi/CFHT) decay of 1E 2259+586 as a function of time (see also Woods et al. 2004 from which the {\it RXTE} data was obtained).}
\end{figure}

\subsubsection{Isolated Neutron Stars}

Half a century since the discovery of the first pulsar,
and with the advent of new radio to TeV facilities, our view of the
neutron star population has changed considerably. Steady and pulsed
emission from isolated neutron stars are detected at all bands, and
despite being allegedly stable clocks, they instead show very
large bursts and flares. In our Galaxy, these flares are second in
brightness only to supernova explosions. Below we will review the
multi-wavelength properties of the magnetar class (the most magnetic neutron
stars; B$\sim10^{14-15}$ G), since they are the neutron star class
that shows the most frequent and energetic transient events. Other
transient events within the isolated neutron star class from which multi-band emission has been
searched (without success thus far) include the Crab
pulsar flares (discovered and detected only in the GeV band), 
pulsar giant pulses (observed mostly in radio), and the fast radio
bursts (detected only in radio so far; although not yet conclusively
associated to young isolated neutron stars or indeed to any other class of source).

\begin{figure}
\begin{center}
\begin{tabular}{l}
        \includegraphics[width=140mm]{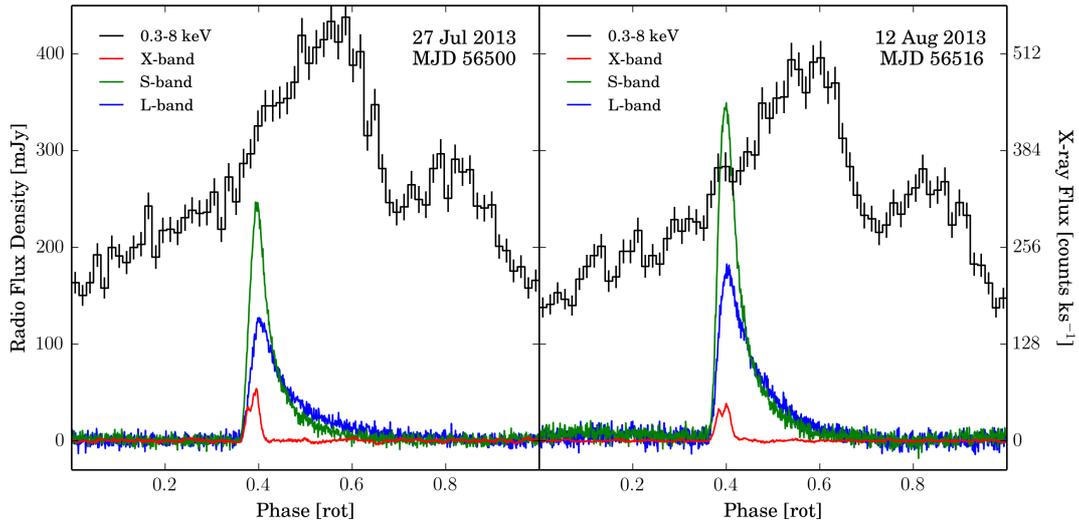} 
\end{tabular}
\end{center}
\vspace{-0.2cm}
\caption{Simultaneous Radio and X-ray pulse profiles of the Galactic center magnetar SGR J1745-2900 (from Pennucci et al. 2015). In order to understand the emission geometry by measuring phase shifts in the pulse-profiles at different wavelengths, strict simultaneity is required.}
\label{fig:l}
\end{figure}

Strongly magnetised neutron stars or {\it magnetars} (recognised in the soft
gamma repeater - SGR - and anomalous X-ray pulsar - AXP - classes) show
large flares at high energies (hard X-rays to soft \g-rays), often confused with short gamma-ray bursts until their
repetition unveiled a different nature (see the reviews of Mereghetti 2008; Rea \&
Esposito 2011; Turolla, Zane \& Watts 2015). Magnetar transient activity may be divided into different
types: a) flares -- usually further subdivided into short, intermediate and giant
flares, depending on the timescales (from ms to minutes),  the
energetics (from $10^{38}$~erg/s up to 10$^{47}$~erg/s), and the
presence of a tail modulated by the pulsar rotation; and b) large
outbursts where the persistent source emission can
increase by several orders of magnitude. Flares and outbursts can be differentiated based on duration, lasting less than one minute for the flares, months-to-years for outbursts, and peak luminosities: between 10$^{38}$ erg/s and 10$^{42}$ erg/s for flares and up to 10$^{36}$ erg/s for the outbursts (for a total energy release during each outburst of
$\sim10^{41-43}$~erg).

Magnetar outbursts are probably caused by large scale rearrangements
of the surface/magnetospheric field, either accompanied or triggered
by plastic motion of the neutron-star crust. These events may result
in renewed magnetospheric activity (through the interaction between
escaping photons from the surface and electrons moving in the strong
magnetic field) and the appearance of additional hot spots on the neutron-star surface, both of which may lead to spectral changes during outbursts,
pulse-profile variability, and different cooling patterns. On the other hand, bursts and flares may (or may not) also be
connected to an internal magnetic field instability that then causes
large magnetospheric rearrangement.

Magnetar flares are so fast and unpredictable that we have no clue
about their multi-band spectrum except in the soft and hard X-ray bands
where most of the all-sky monitors work. In fact, for giant flares -- which are the most energetic and long-lived flaring events -- we lack any simultaneous information about their spectra
below 15~keV. Conversely, the outburst -- lasting months/years -- is more easily followed
in different bands. By following these outbursts, we now know that the
optical and IR emission of these sources are greatly enhanced,
and often follow the X-ray outburst decay evolution (Tam et al. 2004 - Figure 7). On the other hand, in a
few systems, radio pulsed emission is also excited during outburst (Camilo et al. 2006).
Unfortunately, the typically very slow ($>$ day) radio and optical/IR follow-up after an
outburst trigger, and often the limitation in the monitoring
availability in such low energy bands, is hampering our understanding of
the exact physics occurring in these sources.
\bigskip

\noindent {\bf What coordinated observing could reveal:}
\smallskip 

Simultaneous radio-X-ray monitoring of magnetar outbursts might finally reveal the timescale for the radio activation of radio magnetars following the outburst trigger (Rea \& Esposito 2011), shedding light on the connection to normal radio pulsar mechanisms. Currently the activation of the radio pulsed emission following the outburst of a magnetar is constrained to have a minimum delay of about one week after the X-ray activation, and a maximum delay of about a few months, but it may well depend on the particular source and its magnetospheric configuration, as well as on the sensitivity and spanning of the observations. The radio and X-ray observations, if obtained with strict simultaneously, are crucial to understand the emission geometry by measuring any phase shifts in the pulse-profiles in the different energy bands (see Figure 8). On the other hand, IR-optical coverage (with coordination between bands on the timescales of a few days) will instead help address the long standing issue of the presence (or lack thereof) of supernova fossil discs around magnetars (Turolla, Zane \& Watts 2015), and the possible synchrotron or re-processed nature of the IR and optical emission.

\begin{figure}
\begin{center}
\begin{tabular}{l}
        \includegraphics[width=80mm]{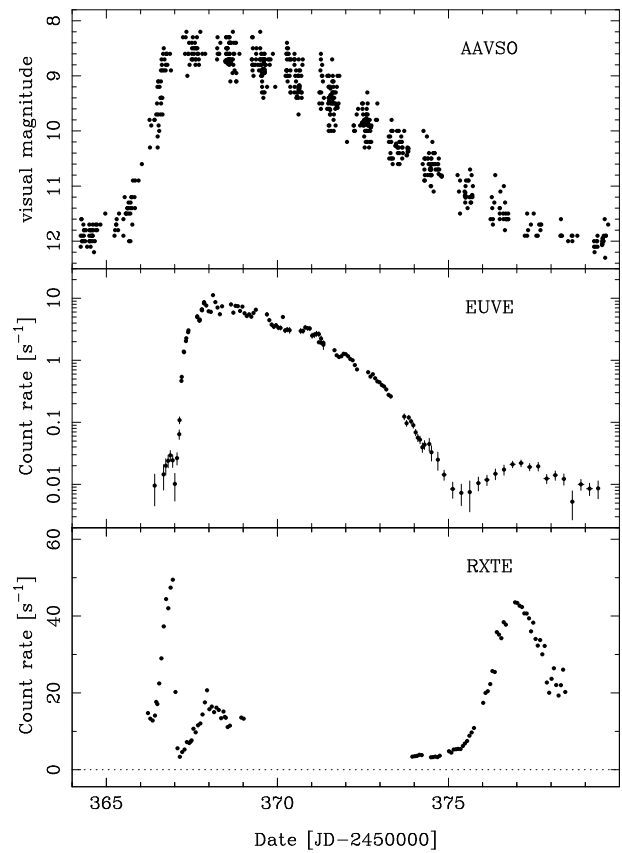} 
\end{tabular}
\end{center}
\vspace{-0.2cm}
\caption{From Wheatley et al. (2003): simultaneous AAVSO, EUVE and {\it RXTE} observations of SS Cygni throughout outburst; although taken over a decade ago, this remains one of the best examples of a multi-wavelength dataset for a DN outburst.}
\label{fig:l}
\end{figure}

\subsubsection{White dwarf binaries}

Accreting white dwarfs (AWDs) are ideal accretion laboratories as they are numerous, bright and harbour a wide variety of accretion flow geometries. In addition, they do not suffer from strong relativistic effects and are therefore not as complex as BH or NSXRBs. AWDs also emit across an incredibly broad energy range from the radio through to \g-rays with their key system components emitting most of their radiation in distinct wavebands. The last point is critical: it means that 
multi-wavelength observations allow us to isolate and {\em dissect}
the accretion flows in these systems. However, all AWDs are variable (showing flickering, orbital variations, superhumps etc.),
and the majority are transients (undergoing dwarf nova and nova outbursts).

The most common type of transient AWD is the dwarf nova/novae (DN/DNe) which exhibit quasi-regular outbursts (recurring on timescales of weeks to decades), during which
their visual brightness increases by factors of $\simeq 10 -
1000$ and the extreme (E)UV light curves (which trace conditions in the hot, inner
disc) tend to lag optical ones (which trace conditions in the outer
disc) by $\simeq 1~$day during the rise to outburst (Lasota 2001).

The interface between the accretion disc and white dwarf surface is thought
to be a geometrically thin {\em boundary layer} (BL) which, in quiescence is hot ($T \simeq 10^8$~K) and produces mainly
X-rays with $kT \simeq 10$~keV. During outburst, the accretion rate onto
the white dwarf increases dramatically and leads to an increase in both the optical depth and
luminosity of the BL which cools more efficiently and emits in the EUV. Thus a picture emerges where near the peak of a DN outburst, there should be an X-ray spike, which is terminated abruptly
once the BL becomes optically thick and starts to produce blackbody-like
emission in the EUV band. 

Empirical tests of this picture require simultaneous X-ray and EUV
observations during the rise of a DN outburst yet very few data sets of
this type exist. The pioneering study of the proto-typical DN SS Cygni
by Wheatley et al. (2003) is probably the best example (Figure 9). In this system, the X-ray and EUV evolution seems broadly
consistent with the standard picture. However, there are also some
clear problems; firstly, in at least some systems, the X-ray flux is
{\em not} quenched during outburst (Mattei, Mauche \& Wheatley 2000;
Byckling et al. 2009). Second, in essentially {\em all} DNe, the X-ray
flux in {\em quiescence} is much higher -- by factors of 100 - 10,000
-- than predicted by the `disc instability model' (DIM - see the review of Lasota 2001). This can be
understood if the inner disc is truncated in quiescence (Schreiber,
Hameury \& Lasota 2003). If so, the difference between the optically
thin ``boundary layer'' in AWDs and the ``corona'' in BH/NS X-ray
binaries might be largely semantic.

As is the case for the eruptions of (BH)LMXBs and (NS)LMXBs, the outbursts
of DNe exhibit clear {\em hysteresis}: the decline from an outburst is
not simply a time-reversed version of the rise. The existence of
hysteretic behaviour in optical light curves has actually been known
for well over 30 years (Bailey 1980; Echevarria \& Jones 1983) and was
quickly shown to be broadly consistent with the DIM (Cannizzo \& Kenyon
1987). Hysteresis effects have also been seen in X-ray light curves of
SS~Cygni (McGowan et al. 2004). However, neither of these examples are
directly analogous to the hysteresis displayed by (BH)LMXBs and (NS)LMXBs in
the HID. The first attempt to construct an equivalent diagram for AWDs
was made by K\"{o}rding et al. (2008). The difficulty here is
identifying spectral bands that track the same components in AWDs as
those tracked by the X-ray bands used in the HID for (BH/NS)LMXBs (i.e. the
inner disc and the corona/BL). As highlighted by K\"{o}rding et
al's HID for the proto-typical AWD, SS~Cygni -- constructed from
X-ray, EUV and optical light curves -- it looks remarkably similar to the
HID of BHXRBs and NSXRBs. However, Hameury et al. (2017) show that 
SS Cygni's HID {\em can} be explained by the DIM, whereas the HIDs of XRBs 
cannot. Whether AWDs also show hysteresis beyond what can be explained by the 
DIM remains an important open question.

Like all disc-accreting astrophysical systems, AWDs appear to drive
powerful outflows in the form of winds, revealed by  the presence of 
blue-shifted absorption and P-Cygni profiles in their UV resonance
lines. These features are not observed in quiescent DNe, where the
resonance lines are always found to be purely in emission and
is reminiscent of the situation in BHXRBs where evidence for disc
winds are mostly detected in the high state (e.g. Ponti et al. 2012).

The other type of outflow seen in many disc-accreting systems are
collimated (radio) jets. Prior to the last decade or so, strongly sub-Eddington AWDs were widely
thought to be incapable of powering jets, with potentially
wide-ranging implications for the physics of jet formation (e.g. Livio
1999). It is worth noting, however, that collimated jets {\em have} been seen in the most luminous AWDs, such as symbiotic stars (e.g. Skopal, Tomov \& Tomova 2013), supersoft sources (e.g. Becker et al. 1998; Motch 1998) and novae during their eruptions (e.g. Sokoloski et al. 2008; Rupen, Miosuzewski \& Sokoloski 2008)

Building on the phenomenological similarities between DN
outbursts and the eruptions of X-ray transients,
K\"{o}rding et al. (2008) obtained radio observations
during the rise of a DN outburst (as this is when (BH/NS)LMXBs tend to show the brightest radio flares). Remarkably, the very
first campaign -- on SS~Cygni -- detected the predicted radio flare just
before the peak of the optical outburst. Since then,
radio emission has also been seen from other DNe in outburst (e.g. Coppejans et al. 2016) but also from some AWDs that accrete
steadily at high rates (Coppejans et al. 2015). 

Although we have focused almost exclusively on DNe - since they provide a clear example of what we have already learned from multi-wavelength observations -- this focus should not be mis-construed as other AWDs also emit across multiple wavebands (and in the context of this white paper are of interest for simultaneous multi-wavelength observations). For
example, {\em novae} -- AWDs undergoing a thermonuclear runaway -- were
only recently discovered to be significant \g-ray sources by
{\it Fermi} (e.g. Cheung et al. 2016), and time-resolved multi-wavelength studies are critical
for understanding the nature and location of the shocks that are likely
responsible for this radiation. Similarly, the unique AWD system,
AE~Aqr, is {\em the} proto-typical example of a magnetic propeller
(e.g. Wynn \& King 1995; Eracleous et al. 1996) and displays extreme,
correlated variability across the entire electromagnetic range, from
radio to \g-rays (e.g. Oruru \& Meintjes 2014). 
\bigskip

\noindent {\bf What coordinated observing could reveal:}
\smallskip 

A huge amount remains to be learnt in general from coordinated long-term observations of AWDs, including the mechanics of the propeller (cf. AE Aqr), the magnitude of accretion disc viscosity, and the nature of the recently-discovered \g-ray emission from novae, to name just a few. Generally speaking, to probe the above physics requires simultaneity on the timescales of tens of seconds to study the flaring in AE Aqr whilst the variability in the disc depends on the viscous timescale and ranges from seconds to hours (depending on the physical properties of the disc and the effective viscosity).

\begin{figure}
    \centering
    \begin{subfigure}[t]{0.45\textwidth}
        \centering
        \includegraphics[width=80mm]{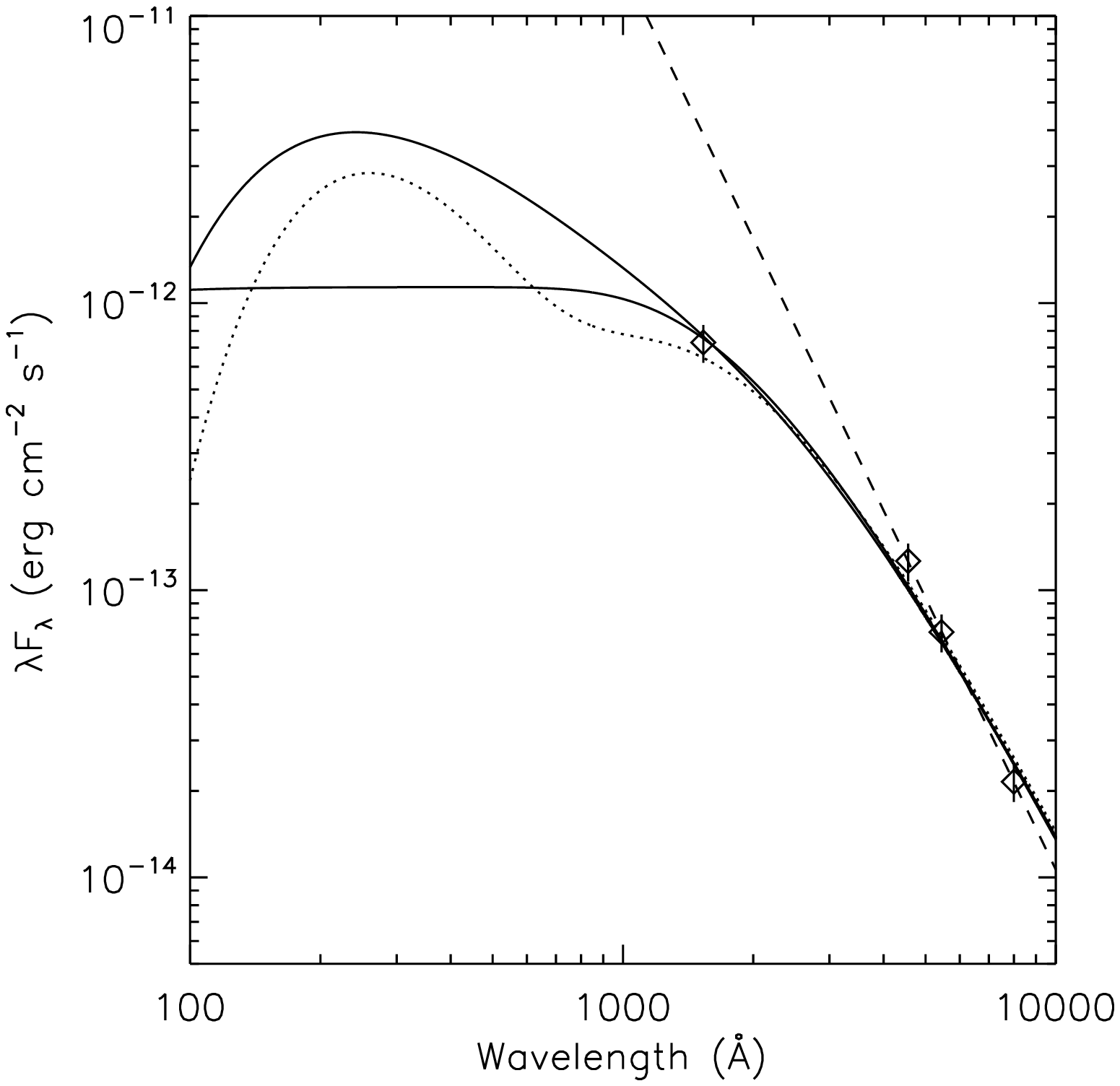} 
        
    \end{subfigure}
    \hfill
    \begin{subfigure}[t]{0.5\textwidth}
        \centering
        \includegraphics[width=80mm]{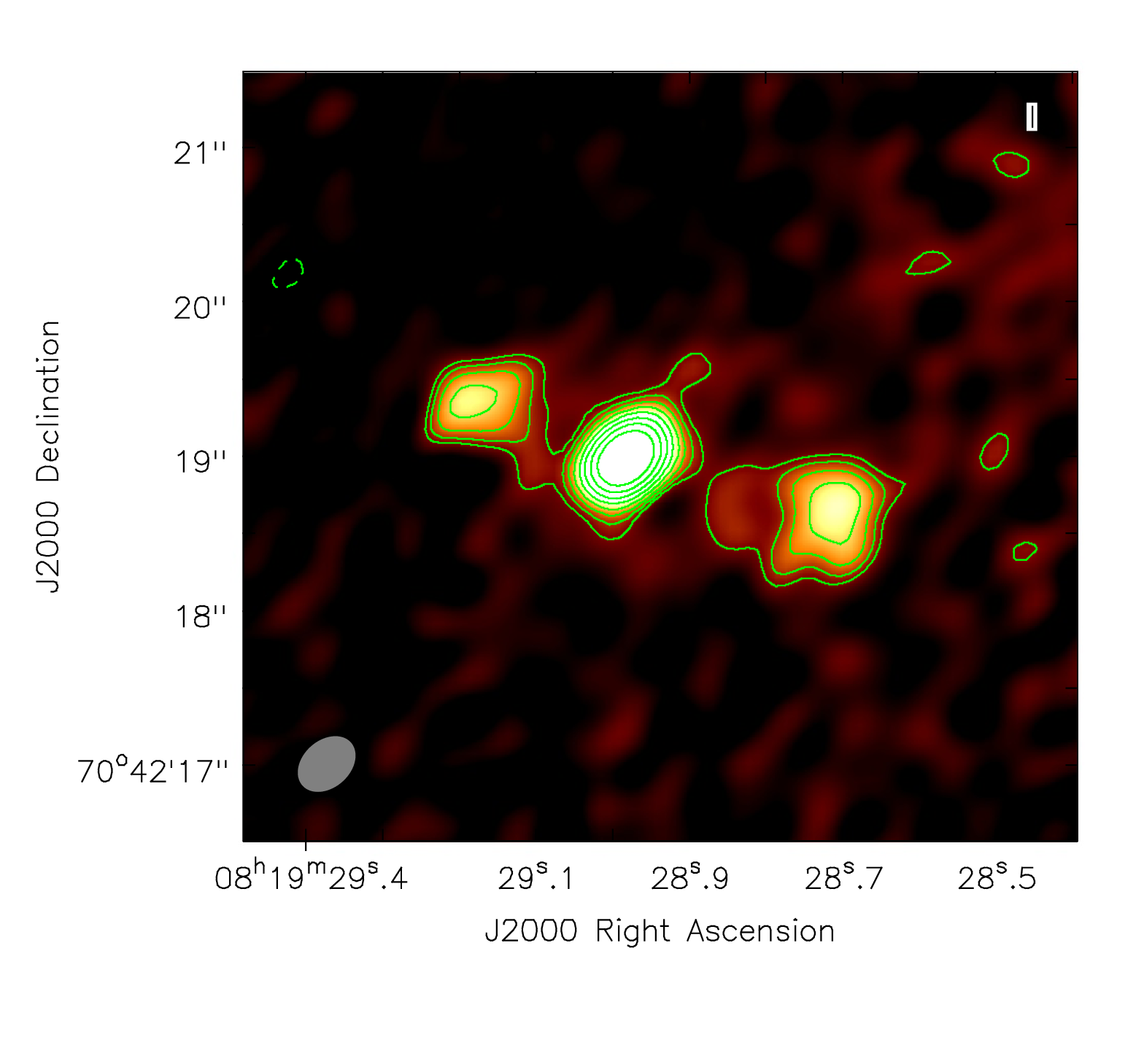} 

    \end{subfigure}
    \caption{Left: UV spectrum of the bright, archetypal ULX NGC 6946 X-1 (with data from {\it HST} and models extending into the X-rays - see Kaaret et al. 2010 from which this figure is sourced). Right: The C-band JVLA A-array image of Ho II X-1 from Cseh et al. (2014) showing the lobe-core-lobe structure seen to evolve in time and indicative of a jet ejection (Cseh et al. 2015).}
\end{figure}

\subsubsection{Ultraluminous X-ray sources}

Since the earliest days of satellite-based X-ray astronomy, extremely bright X-ray sources have been detected in other galaxies but located away from the nucleus and so not identified as AGN. As some of these sources show X-ray luminosities in excess of 10$^{39}$ erg/s - approximately the Eddington limit for a stellar mass black hole around the peak of the Galactic mass distribution (7.8 M$_{\odot}$: {\"O}zel et al. 2010) for an ionised hydrogen accretion disc - they have become known as ultraluminous X-ray sources (ULXs: see the reviews of Roberts 2007; Feng \& Soria 2011; Kaaret, Feng \& Roberts 2017). Due to the difficulties inherent in dynamical mass measurements for such distant sources (where the optical counterpart is faint or often confused), the luminosity (assumed to be isotropic) leads to degenerate scenarios for the mass: either stellar mass remnants accreting at super-critical/super-Eddington rates as seen in the extreme Galactic source SS433 (see Fabrika 2004 for a review) or intermediate mass black holes (IMBHs: M$_{\rm BH} \sim 10^{2-5} M_{\odot}$: Colbert \& Mushotsky 1999) accreting mostly at sub-Eddington rates. 

The discovery of extremely radio-bright ballistic jets in a ULX with L $<$ 3$\times$10$^{39}$ erg/s has demonstrated that those at the faint end are likely associated with accretion onto stellar mass black holes at or close to the Eddington limit (Middleton et al. 2013) whilst those at higher luminosities have unusual spectra (e.g. Stobbart et al. 2006; Gladstone et al. 2009) that can be described as `soft' or `hard ultraluminous' (Sutton, Roberts \& Middleton 2013).  The `low luminosity ULXs' are likely to be one of our best resources for investigating the launching of ballistic jets in BHXRBs as the low absorption column coupled with the well constrained distance to the host galaxy leads to useful constraints on the spin via continuum fitting techniques (e.g. McClintock et al. 2006; Steiner et al. 2014, 2016; Middleton, Miller-Jones \& Fender 2014; Middleton 2015). The bright ULXs on the other hand are likely to be our best option for searching for IMBHs or super-critical accretion. 

Most recently, a dynamical mass constraint for the compact object in a bright ULX has indicated that the unusual spectra of ULXs are indeed due to super-critical accretion rates (Motch et al. 2014). The picture of ULXs as super-critical accretors has been further supported by the discovery of relativistic winds in archetypal ULXs (Middleton et al. 2014; 2015; Pinto, Middleton \& Fabian 2016; Walton et al. 2016) consistent with classical super-critical disc theory (Shakura \& Sunyaev 1973; Poutanen et al. 2007), where the inflow is inflated to large scale-heights, and winds - launched due to radiation pressure - form a conical geometry (see also Jiang, Stone \& Davis 2014; S{\c a}dowski et al. 2014). However, there are stand-out sources whose properties do not obviously fit with those of the wider population, notably HLX-1 and M82 X-2. HLX-1 (hyperluminous X-ray source 1: Farrell et al. 2009) is located in the galaxy ESO 243-49 and shows regular (although non-periodic) outbursts. Such outbursts can reach X-ray luminosities up to 10$^{42}$ erg/s with the source showing spectra that resemble those of (BH)LMXBs (Servillat et al. 2011) although with a much cooler thermal component (and thus a larger inferred black hole mass) and radio emission that could be associated with compact jet emission (Webb et al. 2012; Cseh et al 2015). Studies of the X-ray spectra strongly indicate that the source is an IMBH (see Davis et al. 2011) with optical studies indicating that its presence in ESO 243-49 may have been the result of tidal stripping of a dwarf galaxy or possibly formed in situ within a stellar cluster (Farrell et al. 2012). M82 X-2 shows spectra which appear distinctly harder than those of other bright ULXs (Brightman et al. 2015) and crucially was the first ULX found to harbour pulses in its X-ray lightcurve that indicate the compact object is a neutron star accreting at orders of magnitude above its Eddington limit (Bachetti et al. 2014, which can be somewhat relaxed in the presence of a strong magnetic field, e.g. Mushtukov et al. 2015). Given the recent discovery that two other ULXs also harbour pulsating neutron stars (Israel et al. 2016a, b; Fuerst et al. 2016) with spectra which appear somewhat closer in resemblance to the wider ULX population (see Walton et al. 2017), it is unclear how homogenous the population truly is (see Kluzniak \& Lasota 2015; King, Lasota \& Kluzniak 2017; Middleton \& King 2017)

Although famed for their X-ray brightness, multi-wavelength studies of the brighter ULXs are providing new diagnostics. An archetypal, bright ULX, NGC 6946 X-1 has been found (at least during one epoch) to be ultraluminous (with a luminosity $>$ 10$^{39}$ erg/s) in the UV band (Figure 10; Kaaret et al. 2010; and is consistent with emission from the outer photosphere of a super-critical wind: Poutanen et al. 2007) and another such ULX, Ho II X-1 has been found to launch ballistic jets as revealed by resolved radio emission which change in brightness over time (Cseh et al. 2014; 2015, Figure 10). Finally, several ULXs have been found to be located within large bubble nebulae (tens to hundreds of pcs across) which emit in the optical (and sometimes radio) due to collisional excitation and/or ionisation by the central X-ray source (e.g. Pakull \& Mirioni 2003). Such nebulae provide calorimeters of the long timescale energy output of the source and can constrain the levels of geometrical beaming by the wind cone (King 2009).

\bigskip

\noindent {\bf What coordinated observing could reveal:}
\smallskip 

The coupled multi-wavelength properties of those ULXs which appear to be the high luminosity tail of the BHXRB population (Middleton et al. 2013; Soria et al. 2013) should allow a cleaner view of the coupling between the disc and ballistic jet (i.e. the physics of the jet launching) than can be typically obtained in Galactic sources (by virtue of the lower neutral absorbing column along the line-of-sight). This requires observations in the X-rays and radio (low frequency through to IR) on the viscous timescales of the inner regions (which in turn depends on the prescription underlying the viscosity) and, based on observations of recurrent flaring activity (Middleton et al. 2013) is expected to require simultaneity on sub hour timescales. 

\subsection{Stars: going out with a bang}

\begin{figure}
\begin{center}
\begin{tabular}{l}
        \includegraphics[width=160mm]{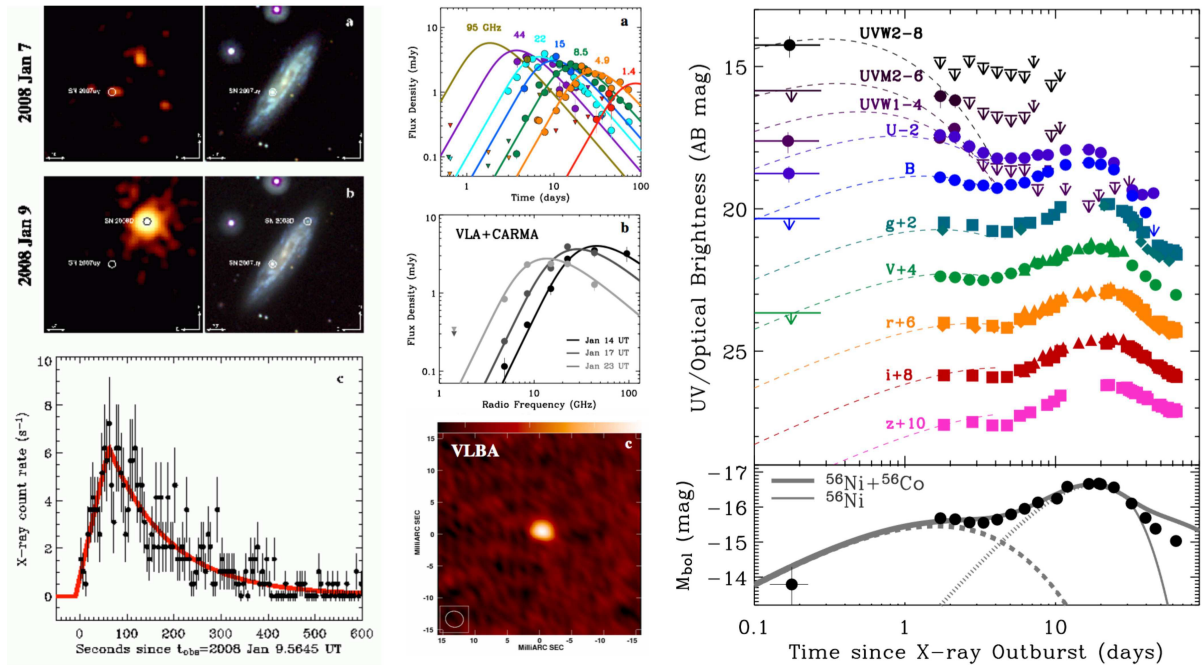} 
\end{tabular}
\end{center}
\vspace{-0.2cm}
    \caption{
    Multi-wavelength observations of SN 2008D from Soderberg et al. (2008), which was discovered
serendipitously via its shock break-out in the X-ray band.
Left panel: X-ray (left) and UV (right) images of the field
surrounding SN 2008D before the explosion on 2008 Jan 7 (a) and during the
shock break-out on 2008 Jan 9 (b) obtained by {\it Swift}. The shock break-out
of SN 2008D is clearly detected during the Jan 9 2008 observations, while
coincident emission is absent 2 days prior. (c) {\it Swift} X-ray light-curve of
the SN 2008D shock break-out showing a fast rise and exponential
decay within 10 minutes of the explosion onset. Middle panel: (a) Radio
light curves of SN 2008D between 1.4 to 95 GHz. (b) Radio spectra of SN
2008D at 3 different epochs. (c) Radio interferometric observation of SN
2008D, placing a physical radius of $\lesssim 2.4 \times 10^{17}$ cm on
the source size. Right panel: (upper panel) The various optical light curves of SN 2008D.
(lower panel) The absolute bolometric (host extinction corrected) light curve of SN
2008D showing possible contributions from radioactive decay (short-dashed
curve) and a cooling envelope blackbody emission model (dashed curve).
Combined models are shown in grey. The optical and radio light curves of
SN 2008D show that observations began within hours of the shock break-out
detection and were obtained quasi-simultaneously out to nearly 100 days
post-explosion.}
\end{figure}

\subsubsection{Supernovae}
 
 Supernovae (SNe) were one of the very first types of astronomical transient ever to be observed by humanity. Even today, studies of ancient texts find references to ``visiting stars", indicating two thousand years of discovery (Green \& Stephenson 2003). The majority of optical light observed from these events comes from the radioactive decay of heavy elements, with the peak brightness scaling with the mass of $^{56}$Ni produced in the explosion (Truran et al. 1967; Colgate \& McKee 1969). From this peak brightness, the ejected mass and kinetic energy of the supernova (SN) explosion can be calculated (Arnett 1980, 1982). As they were discovered in the optical, much of their initial study was performed in this wave-band, with SN classifications being based on the presence or absence of certain types of spectral lines. These optical characteristics imply that there are two different types of SN: thermonuclear (Type Ia) and core collapse. 
\bigskip
 
\noindent {\bf Thermonuclear (Type Ia) supernovae}:
\smallskip
 
Type Ia SNe (SNe Ia) show no H or He lines in their optical spectra, but
have strong Si II absorption lines (Filippenko 1997). Given their uniform
temporal and spectral evolution, along with bright optical luminosities
(absolute \textit{B-}band magnitude of $M_{B} \approx -19.5$), they are
considered ``standard candles", making them ideal distance indicators and
therefore cosmological probes (Riess et al. 1998; Perlmutter et al. 1999).
Recent observations suggest two channels for generating SN Ia. The first
is known as the single-degenerate (SD) model, which suggests the SN is the
thermonuclear explosion resulting from a white dwarf that, through
accretion from a non-degenerate companion, has reached its Chandrasekhar
mass limit. The other scenario is the double-degenerate (DD) model, which
is the merger of two white dwarfs (see Wang \& Han 2012 for a review on SN
Ia progenitor models).

Much has been learned through early-time (within a few days) and late-time
(100s to 1000s days post-explosion) optical observations of SNe Ia. For
example, the detection of a UV flash $<4$ days after the explosion of
SN iPTF14atg, and excess blue light from SN 2012cg at 15 and 16 days
before maximum light, are likely indicative of the SN ejecta colliding
with their non-degenerate companion star, thus supporting a SD explanation
(Cao et al. 2015; Marion et al. 2016). Optical spectroscopic observations
designed to monitor spectral line evolution, taken between a few days
before and up to several hundred days following maximum-light, show
evidence for the SN blast wave interacting with a surrounding
circumstellar medium (CSM) generated by material expelled by the progenitor star during its lifetime. This was the
case for SN 2006X, where the progenitor white dwarf appeared to have been
accreting from a red giant star (Patat et al. 2007). In fact, at least
20\% of SNe Ia in spiral galaxies show CSM interactions, supporting the SD
model for at least some fraction of SNe Ia (Sternberg et al. 2011). Alternatively, optical photometric observations obtained with the Kepler
mission, which have captured several SNe Ia immediately following their
birth, show a lack of early-time (hours to days) brightening that would be
caused by the SN ejecta interacting with a companion star or circumstellar
debris (e.g. Cao et al. 2015; Marion et al. 2016), supporting the DD
scenario (Olling et al. 2015). Additionally, optical observations at
late-times, 100s to 1000s of days post-explosion, when the SN has
sufficiently faded to expose a non-degenerate secondary, have ruled out
the existence of luminous companion (donor) stars. For example, such
observations ruled out a hydrogen-rich (main-sequence) donor star for SN
2011fe, supporting the conclusion that this event followed the DD channel
(Graham et al. 2015; Olling et al. 2015).

Unfortunately radio and X-ray detections of SNe Ia have so far alluded us.
SNe Ia that show optical evidence for CSM interactions are the best
candidates for radio and X-ray counterparts given these wavelengths
usually trace shock-interactions with the CSM (which is the case for
core-collapse SNe; Chevalier 1982). Such observations allow us
to constrain the density of the surrounding medium (e.g. Hancock et al.
2011; Chomiuk et al. 2012, 2016; Margutti et
al. 2012, 2014; P{\'e}rez-Torres et al. 2014); the lack of radio and X-ray detections of CSM interacting
SNe Ia is therefore quite perplexing as this indicates very low density CSM
environments, discouraging some SD models in support of the DD models. For
example, the blue excess detected from SN 2012cg suggests this SN Ia had a
main-sequence companion of $\sim6\mathrm{M}_{\odot}$ (Marion et al. 2016), however, radio observations of this same event provide one of the deepest
radio limits of a SN Ia, suggesting a CSM density that is inconsistent
with most non-degenerate companions, and therefore the majority of SD
scenarios (Chomiuk et al. 2016).
\bigskip

\noindent {\bf Core collapse supernovae}:
\smallskip

Core collapse SNe (CC SNe) result from the core collapse of massive stars. The resulting SN type, which is predominantly based on the existence or absence of H lines, is a direct result of the different amounts of mass-loss the progenitor underwent before explosion (see Smith 2014 for a review). The most common type of SNe are Type II (SN II), which show H in their spectra and come in many flavours based on other spectral characteristics (see Smith 2014 for a summary of subtypes). Such SNe are likely to be the final evolutionary phase of red supergiants with initial masses between $8-20~M_{\odot}$. Type Ib and Ic SNe (SN Ib/c) show no H (and in the latter case no He) in their optical spectra, indicating that prior to the explosion, the progenitor stars underwent mass loss that removed the majority, if not all, of their outer H (and He) envelopes. Such SNe are therefore often referred to as ``stripped-envelope" SNe. Historically, massive progenitor stars of CC SNe have been assumed to evolve in isolation, implying the main mode of mass-loss is via line-driven winds, which are highly dependent on metallicity. While this may be a reasonable scenario for some SN II, it does not explain the extreme mass-loss experienced by the progenitors of SN Ib/c. Recent studies have shown that other mechanisms involving binary evolution through Roche-lobe overflow, or large eruptions of stellar material (like those seen from luminous blue variable stars: Smith, Mauerhan, \& Prieto 2014), may play a crucial role in mass-loss and the final SN (Smith 2014).

In the case of CC SNe, the interaction between the SN shock-wave and the surrounding CSM gives rise to non-thermal radio and X-ray radiation (Chevalier 1982). This non-thermal radiation traces the shock front of the SN, which is the fastest (and in some cases relativistic: Soderberg et al. 2010) moving material ejected by the SN explosion. As the SN shock-wave impacts the dense CSM that was ionised by the initial SN optical/UV/X-ray flash, radio synchrotron emission is generated. The radio emission is initially absorbed, which is usually due to either synchrotron self-absorption from an internal medium or free-free absorption from the external CSM. The flux then rises following a power-law relationship as the density of absorbing material, either internally or along the line-of-sight, decreases until it becomes optically thin (Chevalier 1982; van Dyk et al. 1994; Chevalier 1998; Weiler et al. 2002, 2007). Modelling this light curve behaviour allows us to derive the physical properties of the shock-wave such as the shock-wave radius, magnetic field, average blast wave velocity, and total internal energy of the radio emitting SN material (Chevalier 1998; Weiler et al. 1986; van Dyk et al. 1994). Similar SN shock properties can also be independently calculated from the X-ray emission. Several X-ray emission mechanisms have been suggested for SNe including inverse Compton (IC) scattering, thermal, synchrotron, or central engine driven (see Chevalier \& Fransson 2006 for a review) but the most likely source is IC, where photospheric photons have been up-scattered by a population of relativistic electrons accelerated in the shock front (Bj\"{o}rnsson \& Fransson 2004). The derived shock-wave properties can then be used to directly probe the mass-loss history of the star by calculating the wind speed and mass-loss rate, revealing the final evolutionary stages of the progenitor (Chevalier 1998; Chevalier et al. 2006; Chevalier \& Fransson 2006). 

Extreme forms of CC SNe have now been identified as the source of long duration gamma-ray bursts (GRBs, see the following sub-section). This connection was first demonstrated by the detection of the extremely bright and energetic SN 1998bw that was positionally coincident with GRB 980425 (Galama et al. 1998). Further SN-GRB associations were then made with the strongest link being between SN 2003dh and GRB 030329 (Stanek et al. 2003). Late time ``bumps" in the optical light curves of GRBs were also identified as the rising optical emission from the associated SN (see Woosley \& Bloom 2006, for a review of the SN-GRB connection). These associated SNe had similar characteristics to one another, revealing a new class of SN Ic with broad spectral lines indicative of high velocity ejecta (SN Ic-BL). These SNe associated with GRBs also became known as ``hypernovae" as they are extremely energetic (ejecta kinetic energies $\gtrsim10^{52}$erg) and are likely powered by a central engine (Iwamoto et al. 1998). As mentioned above, radio emission from SNe trace the very fastest moving ejecta from the explosion. The radio observations of SN 1998bw revealed that the ejecta had been moving at relativistic speeds, much faster than that measured from normal SN Ic (Kulkarni et al. 1998). There are also many SN Ic-BL that are not associated with GRBs; radio observations of such objects show that they are unlikely to be powered by a central engine (e.g. SN 2002ap; Berger et al. 2002). In fact, recent studies have shown that the optical spectral properties of SN Ic-BL with and without GRBs are slightly different, with those associated with GRBs having higher absorption velocities and broader line widths (Modjaz et al. 2015). However, there are a few exceptions. Radio observations of the SN Ic-BL, SN 2009bb revealed an engine-driven (relativistic) event similar to SN 1998bw but with no (detected) associated \g-ray emission (Soderberg et al. 2010). A similar case also exists for type Ic SN 2012ap (Chakraborti et al. 2015). One possible scenario is that such an event may be the SN from an off-axis GRB, where the \g-ray jets were orientated away from Earth. Radio searches of other normal, optically selected SN Ib/c without \g-ray counterparts do not reveal other relativistic events, or radio emission associated with off-axis \g-ray jets that have spread into our line-of-sight, implying that engine-driven SNe are very rare (Soderberg et al. 2006, 2010).
\bigskip

\noindent {\bf What coordinated observing could reveal:}
\smallskip

By using (particularly wide-field) telescopes capable of automatically and rapidly responding to transient alerts reported by high-energy, neutrino and gravitational wave facilities (e.g. Kashiyama et al., 2013; Abbott et al., 2016), it will become possible to capture the SN shock exploding through the progenitor star, which is known as a``SN shock break-out". As such an event occurs within minutes to hours, they have only ever been detected serendipitously (e.g. Soderberg et al. 2008 - Figure 11; Schawinski et al. 2008; Gezari et al. 2010; Ofek et al. 2010). Rapid-response, multi-wavelength observations will allow us to detect this phenomenon whilst strictly simultaneous observations at the very early-time (within a few days) will allow us to explore the explosion physics, in turn shedding light on SN progenitors (e.g. Wang et al. 2007; Maeda et al. 2013). 

With specific regards to Type 1a SN, in order to identify the likely progenitor model for a given SN Ia, very
sensitive observations at radio, optical and X-ray wavelengths are
required, with epoch coverage ranging from a few days to 1-3 years
post-explosion. Given the low density of the CSM environments surrounding
SN Ia, it is best to concentrate such multi-wavelength campaigns on the
nearest events (within $20-30$ Mpc). While strictly simultaneous
multi-wavelength observations are not necessary, reasonably contemporaneous
monitoring of SNe Ia at different wavelengths over multiple epochs is
necessary to identify the signatures indicative of an interesting (radio
and/or X-ray bright) event. With high-cadence multi-wavelength detections,
we can then perform light-curve and SED modelling to
identify the likely progenitor scenario for individual SNe Ia.

With specific regards to CC SNe, quasi-simultaneous (within $\sim$the same week) radio and X-ray observations from days to years following a CC SN are important for calculating the thermal excess to any detectable X-ray emission (this is done by extrapolating the synchrotron spectrum calculated from radio observations up to X-ray frequencies), which can directly constrain the mass of the shocked CSM (e.g. Margutti et al. 2017).

\subsubsection{Gamma-ray bursts}  

Our understanding of gamma-ray bursts (GRBs) has progressed
rapidly over the preceding decades, in particular by the advent of satellites that are
able to rapidly localise GRBs and send this information to ground-based observers
on short timescales (e.g. {\it BeppoSAX}: Piro et al. 1998 and {\it Swift}: Gehrels, Ramirez-Ruiz \& Fox 2009). The resulting observational datasets have led to a crude 
model of GRB physics (for a broad review see e.g. Piran 2004): a catastrophic event releases a large amount of energy, 
and relativistic jets are formed. Internal shocks within the relativistic jets create 
the prompt emission, which is what triggers satellite \g-ray detectors.
The interaction of the relativistic blast wave with the circum-source medium (interstellar medium or
the stellar wind of the progenitor star) creates the rapidly fading afterglow: 
synchrotron emission from a forward shock  (travelling into the circum-source medium) and a 
reverse shock (which travels back into the ejecta). As the afterglow fades, the underlying emission
of an associated isotropic transient (e.g. a core collapse supernova) may become visible (e.g. Woosley
\& Bloom 2007).
After all transient emission is gone, most GRBs show a faint host galaxy (e.g. Hjorth et al. 2012). 

Broadly speaking one can divide GRBs into (at least) two classes, the short-hard and long-soft
GRBs (see Kouveliotou et al. 1993). The long duration bursts are observationally linked to supernovae, whereas the 
short duration bursts are generally thought to arise in compact binary mergers (e.g. NS-NS or 
NS-BH mergers). In both cases the association of a GRB with a progenitor channel is studied 
via prompt emission energetics and host galaxy properties (e.g. offset from star forming regions), 
but a more direct diagnostic are their late-time transient signatures which require multi-wavelength observations to disentangle contributions from host galaxy and afterglow.

In studying the prompt emission, the very short duration (seconds) makes it difficult to obtain multi-wavelength observations. However, the expanded bandpass of triggering instruments (e.g.
{\it Fermi} LAT with {\it Fermi} GBM), the advent of fully robotic ground-based telescopes and the short {\it Swift} slew 
times has meant the wavelength coverage of 
bursts during this phase has increased. In turn, this has allowed more robust time-averaged and time-resolved spectral fitting (e.g. Ackermann et al. 2014). Despite these new observational inputs, the exact nature of 
the prompt emission spectrum is still poorly understood. 

\begin{figure}
\begin{center}
\begin{tabular}{l}
 \epsfxsize=14cm \epsfbox{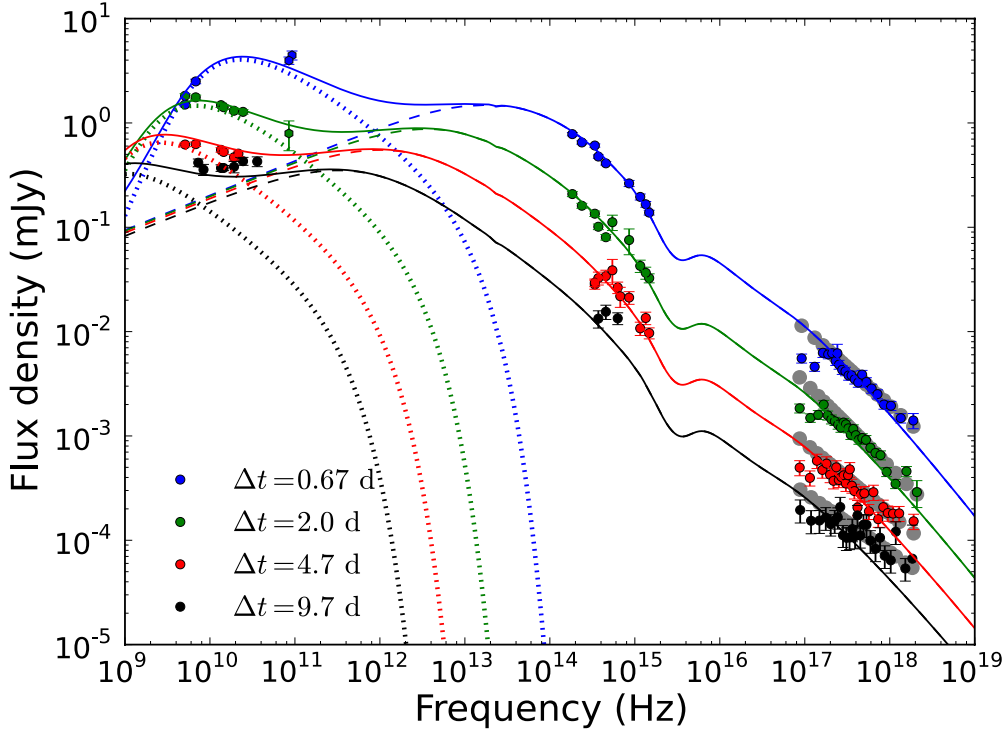}
\end{tabular}
\end{center}
\vspace{-0.2cm}
\caption{SED of the GRB 130427A afterglow covering nine orders of magnitude in frequency and shown at various decay times (taken from Laskar et al. 2013). The spectral contributions from the forward- and reverse-shock are indicated by dashed and dotted curves, with the solid line showing the sum of the two components. Such remarkable multi-wavelength data and temporal sampling is clearly vital to test emission models for these powerful events.}
\label{fig:l}
\end{figure}

\begin{table}
\begin{center}
\begin{minipage}{180mm}
\bigskip
\caption{Radiative processes}
\begin{tabular}{|c|c|c|c|c|c|c|c|}
 
\hline
 Object & Radio & sub-mm & IR & optical & UV & X-ray & \g-ray \\
   \hline
AGN & S & S & rep & T &  T & IC/SSC & IC/SSC \\
(Seyferts) & & & & & & & \\
Blazars & S & S & S & S & S/IC/SSC & S/IC/SSC & IC/SSC\\
Sgr A*  & & & & & & & \\
TDEs & S & S & S & T & S/T/IC & S/IC & IC \\
BHXRBs & S & S & S/*/rep & S/*/rep/SSC & S/*/rep/SSC  & S/T/IC & S/IC \\
NSXRBs & S & S & S/*/rep & */rep/SSC & */rep/SSC & NS/T/B/IC & ? \\
Iso. NS & S & S & S/rep & S/rep & ? & T/RCS/IC  & S/B/RCS \\
AWDs/DNe  & S & ? & T/* & T/* & WD/T & T/B & ? \\
ULXs & S & ? & */S & */T/rep & */T/rep & T/IC & ? \\
SNe & S & S & T & N/T & T & S/T/B/IC & IC/N\\
GRBs & S & S & S & S & S & S/SSC/IC & S/SSC/IC \\

\hline
\end{tabular}
\vspace{0.5cm}

Notes: A broad summary of the possible radiative processes in each object class discussed so far. The key for the processes is as follows: T = optically thick thermal emission (i.e. a blackbody/modified blackbody), S = synchrotron, IC = inverse Compton (including bulk Compton), SSC = synchrotron self Compton, B = bremsstrahlung, rep = reprocessed emission (i.e. a second order thermal process) * = stellar companion, N = nuclear processes, RCS = resonant cyclotron scattering, ? = unknown or no current observations in this band. We note that not all of these processes happen at once, e.g. the prompt emission from GRBs radiates in the X-rays and \g-rays via SSC/IC whilst the afterglow does not (instead radiating in these bands via synchrotron losses).

\end{minipage} 

\end{center}
\end{table}

The afterglow that follows the prompt emission consists of two components, the forward and 
reverse shocks (for example, see Figure 12). The forward shock afterglow is visible over many decades in frequency, from 
radio to \g-rays, and has a simple synchrotron spectrum, the break frequencies of which
(and peak flux) evolve rapidly with time, as the dynamics of the blast wave evolves (e.g. Sari, Piran \& Narayan 1998). Analytical models thereby allow the key micro- and macro-physical parameters of 
the afterglow to be retrieved through SED and light curve fitting (e.g. Wijers \& Galama 1999). The reverse shock travels back into the ejecta, and therefore is bright but short-lived -- but may be one of the 
few ways to probe the ejecta properties, like the magnetisation, electron population and baryon loading. 
Observations with rapidly responding (robotic) telescopes have detected
optical (which can be nearly naked-eye brightness)
 and radio reverse-shock flares (e.g. Japelj et al. 2014 and Anderson et
al. 2014). It is likely that optical detections are relatively rare
because the typical wavelength of a reverse shock is more often located
redwards of the optical range indicating the need for more robotic radio
facilities. Constraints on the occurrence rate and strength of 
reverse shock flashes, and their spectral behaviour in optical and radio wavelengths, strongly constrain the 
physical parameters of the reverse shock and the ejecta, in particular the magnetisation (e.g. Japelj et al. 2014, Granot \& van der Horst 2014).

In the above, we have discussed the basic lightcurve- and SED-based successes coming from multi-wavelength endeavours yet
it is also worth mentioning multi-wavelength polarimetry as an additional tool. Measurements of the polarisation degree and 
position angle as a function of time and at various wavelengths provide additional information
to constrain model sets (e.g. Toma, Ioka \& Nakamura 2008), as an example, \g-ray polarimetry probes the coherence and strength of magnetic fields 
in the internal shocks and diagnoses the emission mechanism (Yonetoku et al. 2012) whilst optical and radio polarimetry of the forward shocks test the microphysics, internal structure and collimation of the jets (e.g. Wiersema et al. 2015).
\bigskip

\noindent {\bf What coordinated observing could reveal:}
\smallskip 

With extensive spectral coverage (radio up to X-ray wavelengths from early to late times) the modelling of multi-wavelength data can be used to constrain the micro- and macro-physical properties of the GRB such as the blast wave kinetic energy, the density of the surrounding medium, the total energy budget, and probe the jet dynamics of these extremely energetic events (see Granot \& van der Horst 2014, and references therein). At early times 
(within a day following the GRB), simultaneity requirements are generally hours or less 
for optical/IR and X-rays, and up to a day or so for longer wavelengths. At later
times following the GRB, these requirements are more relaxed as the characteristic
timescales are slower. 
 
 \subsection{Summary - processes and timescales}

As can be seen from the previous sub-sections, non-simultaneous, multi-wavelength observations have provided new, important insights into a wide variety of sources and physical processes. However, in the case where the emission at different energies varies on timescales {\it shorter} than the coordination between bands, we are provided with at best a fundamentally limited view and at worst a misleading one. The relevant timescales that need to be considered -- which determines the difficulty inherent in coordinating observations -- depends on the physical processes under consideration and the object class. Some of the specific timescales and bandpasses are discussed above, yet it is valuable to consider a very broad view of the overlap of timescales and bandpasses as these play a role in synergy between instruments and fields.

In Table 1 we show a summary of the radiative processes that {\it can} contribute to a given band by a given source but we stress that the process depends on the source state, e.g. the prompt emission from GRBs radiates in the X-rays and \g-rays via SSC/IC whilst the afterglow does not (instead radiating in these bands via synchrotron losses). We have also taken care to include all the possible (debated) origins and as multi-wavelength observing improves, some of these may vanish (whilst others may appear in their place). In Figure  13 we show the subclasses relative to where they emit a sizeable component of their energy in the EM spectrum and the rough correlated timescales they may operate on - which in turn gives some idea of the level of simultaneity required in their study.

 \begin{figure}
\begin{center}
\begin{tabular}{l}
 \epsfxsize=12cm \epsfbox{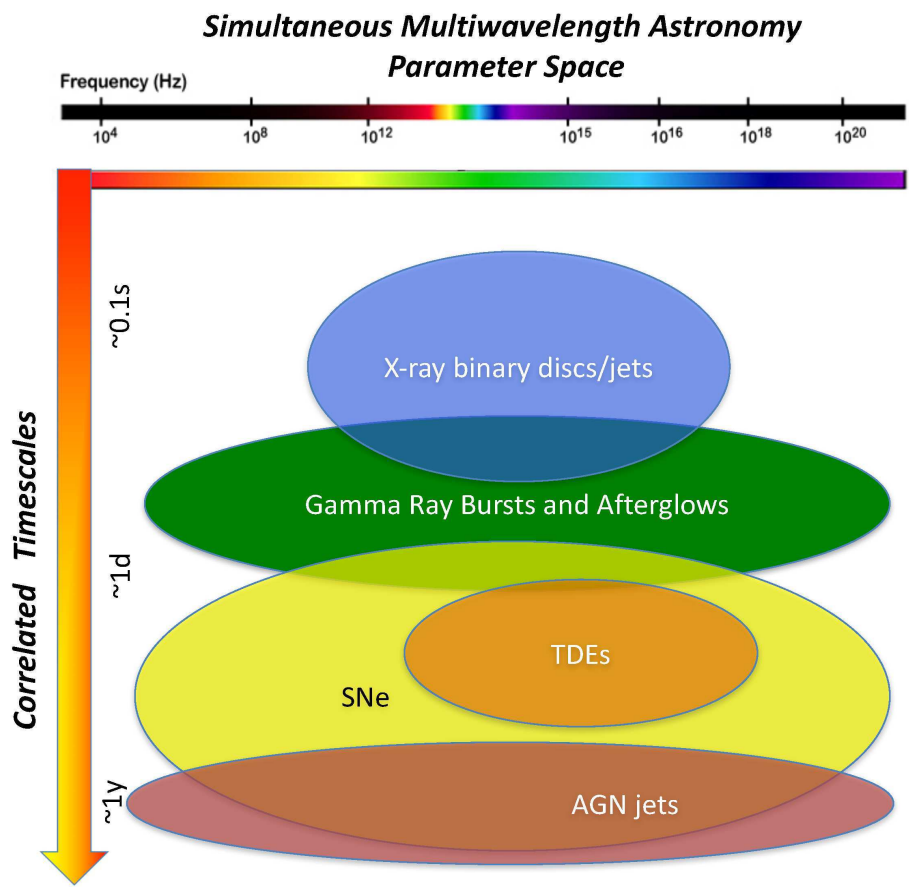}
\end{tabular}
\end{center}
\vspace{-0.2cm}
\caption{Here we provide an indication of the synergy between object classes, their relevant variability timescales and the wavebands over which they emit (the x-axis is contracted in colour to indicate the full 10$^{4}$ to 10$^{20}$ Hz bandpass).}
\label{fig:l}
\end{figure}

\section{Current options for studying transients, synergy and limits on multi-lateral observing}

The difficulty in performing meaningful multi-wavelength observations depends on the target and process under study (see preceding section). Simultaneity within a few days can be acceptable when observing  slowly varying targets at multiple wavelengths (as for example AGN jets), but it is not a very useful option when the objective is, for example, to obtain a broad-band SED of a (BH)LMXB, which can vary its spectral properties over timescales shorter than a day (see Figure 13), forcing observers to obtain data at multiple wavelengths at least during the same night. As indicated in the previous section, for some objects/processes, the relevant timescales can be as short as milliseconds, which clearly sets the bar at a completely different level. The predictability of the event is also a crucial ingredient, not necessarily correlated to the characteristic timescales explored. A campaign aimed at studying the correlated multi-band, sub-second variability of an XRB, will meet very different obstacles than those faced by GRB observers when trying to cover the time evolution of the broad-band spectral emission of a new GRB candidate. While the former requires strict simultaneity, it can be scheduled days in advance, and can in principle last for just a few hours; conversely, GRB observations need to be activated and scheduled with very short notice, can last for several days (to several weeks for the afterglow at radio frequencies) and can often achieve only quasi-simultaneity (within a few hours).

\subsection{Issues faced in coordinated observing} 

Overall, the main difficulties inherent in arranging coordinated campaigns (with no emphasis on simultaneity) can be summarised in the following broad terms:

\begin{itemize}

\item[--] \underline{Obtaining observing time.} A successful multi-wavelength campaign depends on observing time from multiple facilities. Although several facilities presently offer the possibility to submit a `joint proposal', the available combinations are still minimal, and with many limitations, for example, some joint calls do not allow proposals for ToOs, as is the case for the XMM-ESO joint program and some countries have limited or no access to certain facilities. Thus, researchers are forced to apply independently to different observatories, which leaves the door open to multiple jeopardy. In cases when the scientific objective is truly multi-wavelength, the proposals can be (and often are) rejected because time at other wavelengths cannot be guaranteed. Moreover, proposal deadlines are spread over the year, and are valid for different and often not aligned (year-, semester- or trimester-long) observing periods, making the whole procedure extremely difficult.

\item[--] \underline{Scheduling.} Once observing time is secured at all the required wavelengths, successful proposers then face the scheduling challenge: different facilities, each with their manpower limitations, internal timescales, scheduling issues and constraints, etc., need to coordinate their efforts to schedule (potentially simultaneous) observations. The visibility for the observation represents an additional, major challenge, in that it is often left to the PI of the program to identify the windows of common visibility among all the involved facilities. This difficulty is also directly linked to the type of observing campaign, with programs requiring very fast response times facing clearly the most substantial (and at times seemingly insurmountable) obstacles. 

\item[--] \underline{Human limitations.} We cannot work on everything. We cannot take care of everything. Sentences like``you could have asked me, I have an approved program for that" are often heard at conferences. Researchers focus on their own program objectives, within the boundaries of their own (in most cases band-limited) expertise, often unaware of the potential their data might have for other observers, and of the possibilities that other observers' data would offer to them. Very often, researchers do not even know of existing and potentially useful facilities, or even if they do, they do not take them into consideration for their program because of the unknown technical challenges, or because of perceived or genuine lack of access (e.g in cases where the facilities are privately run by single institutions). Additionally, observers often do not know the schedule of many facilities, either because this information is not made available or because of personal lack of expertise in that band.

\item[--] \underline{Rarity.} Another important practical issue is rarity of object class. Many transient systems e.g. SNe, GRBs are relatively common when compared to XRB outbursts (especially those extreme events such as the 2015 outburst of V404 Cygni) and TDEs (at present) and so observations are -- in principle -- much easier to coordinate amongst observatories; indeed it is possible to \lq bank\rq\ ToOs in advance and to attempt coordination on a case-by-case basis. The case of rare transients (TDEs, XRBs in outburst) is, undoubtedly, much more difficult to accommodate, however, it is often possible to decide upon a science strategy in advance, and to choose an appropriate target later on, in near-real-time (e.g. the case of Doppler tomography of Galactic AWDs in outburst). 

\end{itemize}

\subsection{Issues specific to coordinated {\it simultaneous} observing} 

As we have discussed, the issue of simultaneity between observing bands (and its various practical definitions) is relevant for essentially all sources we can study. Naturally, simultaneity has differing thresholds depending upon the science and on transient class (see Figure 13 and the preceding section); e.g. whereas AGN observations may be considered closely simultaneous if coordinated at different wavelengths to within a few days, observations of XRBs or GRBs require up to millisecond levels of simultaneity. We have identified several key practical issues which currently hamper progress with specific relevance to {\it simultaneous} multi-wavelength follow-up which, for the most-part, do not depend on source class (although, as we have already mentioned, the difficulty naturally increases with decreasing timescales): 

\begin{itemize} 

\item Response times to ToO triggers and DDT requests at certain observatories can take a long time to process depending upon available manpower, while some observatories do not consider time-critical observations to be a key driver and so are not prepared to handle such requests.  

\item As with obtaining non-simultaneous multi-wavelength observations, obtaining simultaneous observations can often involve {\it double-jeopardy} with multiple observatories, with one observatory wishing to commit time only after another observatory has agreed to do so. Given that observatory proposal deadlines are often not synchronised with each other, it is often hard to get time approved for simultaneous observations. 

\item Current proposal systems do not even allow for certain types of proposals to be submitted. A particularly relevant example is the lack of a very large multi-observatory ToO campaign, which cannot be accommodated in current systems due to its ambitious nature. 

\item Simultaneous observations are not occurring as often as they could be, because telescope observing schedules are often not known (or cannot be known due to uncertain weather conditions). Whereas some observatories report their short-term, or real-time, observing schedules, most do not. 
  
\end{itemize}

 \begin{figure}
\begin{center}
\begin{tabular}{l}
 \epsfxsize=15cm \epsfbox{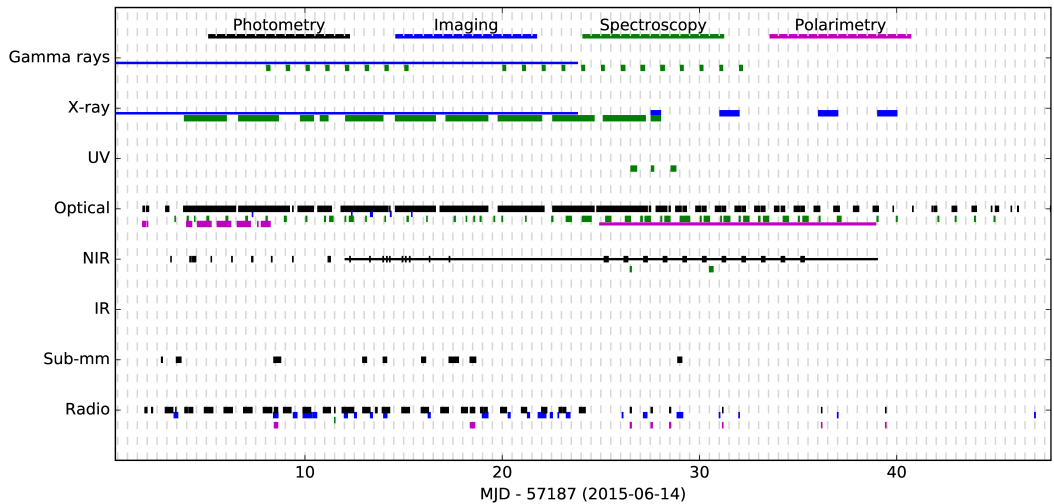}
\end{tabular}
\end{center}
\vspace{-0.2cm}
\caption{A depiction of the large-scale, coordinated, multi-wavelength campaign during the 2015 outburst of V404 Cygni. (credit: Tom Marsh)}
\label{fig:l}
\end{figure}

\section{When coordinated observing works: the case of V404 Cygni}

Formally established collaborations have a long history of being able to coordinate observations at multiple wavelengths, and over long periods of time. The {\it AGN Watch} collaboration is just one example of such a success story (e.g. Peterson et al. 2002, and references therein). For more {\it ad hoc} and real-time coordination amongst observers, the need for rapid dissemination of information has long been recognised; the IAU Central Bureau for Astronomical Telegrams and The Minor Planets Circulars have served this purpose for more than half a century. The advent of electronic communications allowed near-instantaneous communication of results via the internet, through free services such as The Astronomer's Telegram\footnotemark\footnotetext{http://www.astronomerstelegram.org} and Gamma-ray Burst Coordinates Network\footnotemark\footnotetext{http://gcn.gsfc.nasa.gov} (GCN). These facilities are typically used not only for reporting the results of observations of transient events, but also for encouraging and motivating follow-up observations. This model has proven to be highly successful in every field, whether used for localisation of GRB counterparts, determining redshifts of SNe, or in obtaining complete observational coverage of XRB outbursts. 

As one recent example of successful coordination, we can look at the June 2015 outburst of the Galactic (BH)LMXB V404\,Cygni; this was one of the brightest black hole binary outbursts in recent memory, and provided an excellent testbed of the effectiveness of various coordination networks. The source reached peak luminosities rivalling the Eddington luminosity of a 9 M$_{\odot}$ black hole during several flares (Rodriguez et al. 2015). The main phase of the outburst lasted about two weeks, with highly complex patterns of multi-wavelength variability reported across the electromagnetic spectrum, and on all timescales down to $\sim$10 ms (Kimura et al. 2016, Gandhi et al. 2016). 

Notable, widely publicised, real-time coordination efforts were made centred on a {\it HST} observation (Knigge et al. ATel 7735). This not only included a call for coordinated follow-up, but also the creation of a dedicated mailing list ({\tt v404-mwc@lists.soton.ac.uk}), as well as a webtool for easy coordination of observations ({\tt http://deneb.astro.warwick.ac.uk/phsaap/v404cyg/data/}). At peak, the mailing list had over 135 subscribers. The {\it HST} observation was, unfortunately, cancelled as a result of the source fading, but coordinated observations by other observatories at multiple wavelengths did continue (Knigge et al. ATel  7773). 

Near-continuous coverage of the outburst of V404 Cygni was provided in X-rays, radio and  \g-rays over several days by the {\it INTEGRAL} observatory, the arcminute micro-Kelvin imager (AMI), the {\it Chandra X-ray observatory} and {\it Swift}. The {\it INTEGRAL} observation was carried out as a Public ToO observation and was advertised by Kuulkers et al. (2015, ATel 7695), encouraging simultaneous follow-up and allowing their data to be used in several studies (e.g. Kimura et al. 2016, Gandhi et al. 2016). Also at X-ray energies, the {\it Chandra X-ray observatory} provided several DDTs and ToOs which, via the onboard high-resolution gratings, led to the discovery of mass loaded outflows visible in the X-rays (clearly identified via a P-Cygni profile: King et al. 2015) - a key diagnostic for the accretion flow.

At optical wavelengths, the American Association of Variable Star Observers (AAVSO) observed the source almost continuously with an army of small (1--3\,m) and tiny ($<$\,1\,m) telescopes spread over the globe; Kimura et al. (2016) made extensive use of these data to draw inferences about the optical variability patterns seen from the source. In addition, the value of having multiple (quasi)simultaneous observations from various telescopes for cross-calibration purposes cannot be understated (see the extensive discussion in Gandhi et al. 2016). In short, the 2015 outburst of V404 Cygni was undoubtedly one of the most collaborative experiences for the XRB community in terms of multi-wavelength global coordination efforts, and represents an important milestone for the community as a whole. To give an impression of the scale of the coordinated efforts, Figure 14 shows how the different wavebands overlapped across the course of the outburst. 

As opposed to the community efforts to study V404 Cygni, the vast majority of ATel and GCN circulars remain dedicated to quick reporting of results, rather than for specific coordination of future observations. Most coordination still happens by word-of-mouth (be it virtual or face-to-face), and depends on known personal networks of collaboration. In the new era of large-scale transient discovery (see the following section), this approach is likely to prove inefficient for coordinating simultaneous follow-up, and new strategies will be required.

 \begin{figure}
\begin{center}
\begin{tabular}{l}

 \includegraphics[width=90mm, angle=-90]{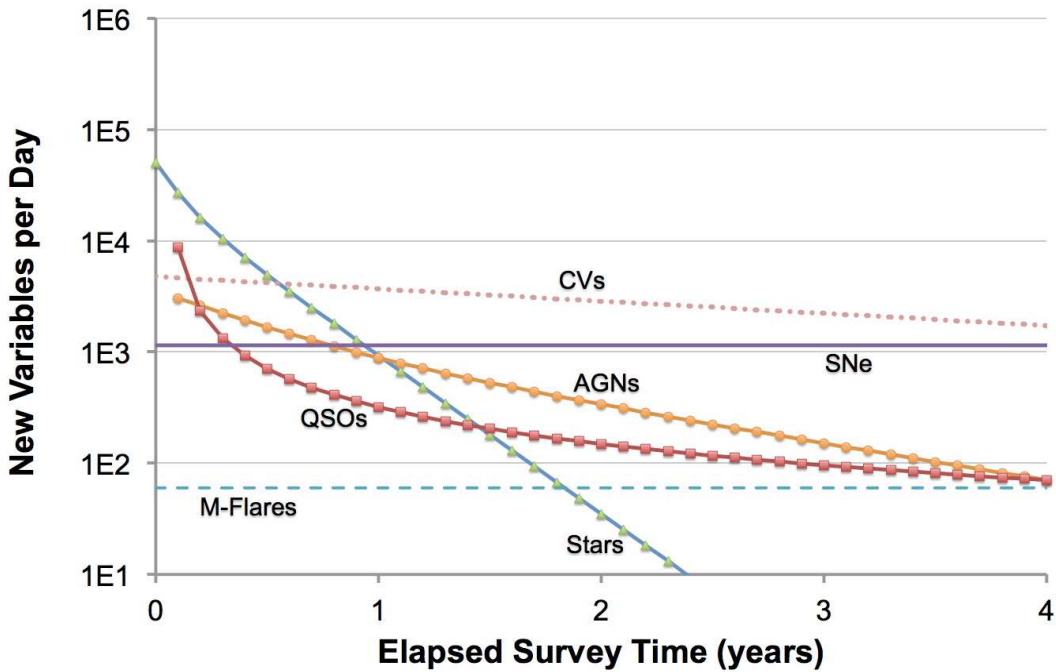}

\end{tabular}
\end{center}
\vspace{-0.2cm}
\caption{Adapted from Ridgway et al. (2014). As an example of the impact of the synoptic era on transient detection, LSST will discover a raft of new transients of all classes (although note that the expected number of detected CVs is an extreme upper limit) - determining how we follow-up such a huge number of sources, i.e. source-selection, coordination etc. is a major challenge we must overcome as a community.}
\label{fig:l}
\end{figure}

\section{Entering the {\it Synoptic era} - the increasing demand for rapid followup and coordination}

We are entering a new era of astronomy, in which large facilities will observe and monitor large fractions of the sky, throwing on the table an immense amount of information. Dense and regular monitoring of a staggering number of targets will be guaranteed at several wavelengths, with millions of potential transients being discovered daily (see Figure 15). This changes drastically each and all of the difficulties listed in the previous section. In some cases, they will be effectively solved by the very existence of these facilities and their fantastic data throughput. Other issues will however become even more pronounced as the numbers of transient sources becomes rapidly so large as to quickly outpace the feasibility of handling data and information via traditional means, and of carrying out certain more specialised follow-up (e.g. spectroscopy, VLBI, polarimetry). 

A crucial issue will be the need to automate most of -- if not the whole -- process: transient detection, first-look classification, prioritisation, communication, trigger of multi-wavelength follow-up and further classification. If this may appear simple to conceptualise, the actual implementation is and will be a tough challenge, involving technical software difficulties, political issues, and scientific challenges. This topic alone would require pages of discussion and analysis, which is beyond the scope of this document and so we limit ourselves to some of the issues that were raised/discussed during the meeting. 

\begin{itemize}

\item[--] \underline{Political issues.} Some of the facilities providing target detections will have strict policies, with formalised Memoranda of Understanding (MoUs) binding collaborators to specific sharing rules. With several facilities joining in and activating more or less automated follow-ups, there can be an issue of how not to reveal pointing coordinates.

\item[--] \underline{Prioritisation.} The saturation limit will be hit very quickly, with thousands of potential triggers arriving from different facilities. Follow-up observations will also provide further triggers, thus hard choices will need to be made in order to decide which events to follow.

\item[--] \underline{Keeping track of what's going on.} There will be many facilities providing different types and amounts of information, all simultaneously. Triggers and follow-ups will pile up in the schedule of telescopes all around the world in a way that will be difficult to follow individually. There will therefore be an urgent need for `information sharing points' or brokers to collect all the information and make them available to the whole community in an easy format.

\item[--] \underline{Observing philosophy.} We need to foresee a fraction of `randomly selected' follow-ups, to leave the door open to the unexpected (see the following section).

\end{itemize}

\section{Serendipity and the impact of Gravitational Waves and Fast Radio Bursts}

It is important to emphasise the role serendipity plays in astronomy; whereas most observatories and large missions are designed with some key scientific goals, it is often the case that the lasting impact of these missions is something unexpected -- discoveries that were not anticipated at inception. Examples include the discovery of gamma-ray bursts, the cosmic X-ray background radiation, and accelerating universe. Therefore, there is much to be said for leaving the door ajar and being prepared for chance discoveries (e.g. Fabian 2009). With increasingly sensitive and wide-field instrumentation across the electromagnetic spectrum, presumably yet undiscovered classes of object will be revealed in coming years; as shown by the history of GRBs, multi-wavelength observational strategies are likely to be required to identify these objects and probe their physics in detail.

Future discoveries aside, examples of new classes of objects and new physics currently exist. The discovery of gravitational waves (Abbot et al. 2016a) from merging black holes recently opened a rich new window into the study of astrophysics, bringing to a close a century of prediction and search. The detection of the electromagnetic signatures of astrophysical events that produce gravitational waves is already an impressively coordinated activity, as shown in the electromagnetic follow-up of GW150914 (e.g. Abbott et al. 2016b) that returned a null result (expected from a binary black hole merger). Detecting the electromagnetic signature of a gravitational wave event (likely involving a compact object that is not a black hole) would allow localisation of the event, possible identification of a host galaxy, and detailed investigation of the energetics of the event.  

Given the importance of gravitational waves to fundamental physics (not just astrophysics), there is strong motivation to coordinate multi-wavelength observations that cover large sky areas to match the very poor localisation capabilities of the current gravitational wave detector network.  The level of coordination for gravitational wave follow-up is exactly the same as for many other classes of objects that are interesting in the time domain.  In fact, from campaigns such as those described by Abbott et al. (2016b), it is possible that a wealth of information could be obtained for the very many AGN, galaxies, and stellar objects that find themselves in the gravitational wave error region and subject to multi-wavelength observations.  Thus, possibilities for commensal science based on gravitational wave follow-up should be given some thought.

Another exciting example of a new class of objects of importance is Fast Radio Bursts (FRBs: Lorimer et al. 2007), millisecond timescale bursts of radio waves that have a large frequency dispersion sweep.  Whereas gravitational waves have been predicted for a century, a decade ago FRBs were discovered without any theoretical prediction.  Whilst FRBs have remained enigmatic, as more are detected in the radio band, it is becoming reasonably clear that they have an extragalactic origin.  Thus, the large dispersion presumably reflects the large electron column the signals traverse before detection on Earth, potentially giving an unprecedented probe of the intergalactic medium (IGM) on cosmological scales (e.g. Macquart et al. 2015).  

Now that FRBs have been shown to be highly interesting, the motivation to coordinate multi-wavelength observations has grown rapidly.  Much of this motivation is similar to that for gravitational wave follow-up, identifying host galaxies in order to measure redshifts and distances, to allow a probe of the IGM via the dispersion and other properties of the FRB signals.  An early demonstration of such a strategy was for FRB140514 (Petroff et al. 2015).  A later effort by Keane et al. (2016) claims success in identifying an afterglow for FRB150418 at radio frequencies.  This result remains controversial, as further work indicates the presence of a variable AGN in the claimed host galaxy (Berger \& Williams 2016; Bassa et al. 2016).  FRB150418 shows that not only multi-wavelength coordination but multi-scale resolution can sometimes be required.

The recent discovery of the first repeating FRB by Spitler et al. (2016) provides strong hope that it can be localised and a progenitor/host can be identified, since the error region is small and concerted observational efforts can be made;  while FRBs were only single pulses from an unpredictable point on the sky, coordinated observations were extremely difficult.  However, even with identification of the repeating FRB, it is unclear if this object is representative of other FRBs, as no other has ever been seen to repeat.  Ongoing multi-wavelength efforts will no doubt be required to better understand the FRB population and unlock their potential as astrophysical/cosmological probes.

With examples such as FRBs and gravitational wave events as prime physics targets for future coordinated multi-wavelength observing campaigns, these seem like good places to start when defining a cross-facility strategy.  The requirements in both cases are almost identical to the requirements for other classes of time variable objects of interest.  As high impact areas of research, observatory directors are likely to be more motivated to coordinate (as already demonstrated in the case of gravitational wave case) and `hitching a lift' on this coordination to achieve additional astrophysics for other object classes is worthy of consideration.

\section{What needs to be done}

The workshop leading to this white paper included a number of strategy sessions dedicated to the requisite steps to improve the state of coordination. Many (but not all) of the identified problems can only be solved by policy changes at a director level, however, we hope that the solutions we arrived at at the workshop and present below will be a useful starting point for discussions and will aid in the lobbying of those with the power to make the necessary changes. 

\begin{itemize}

\item[--] \underline{Scheduling of present instruments.} A possible means to avoid the issue of double-jeopardy, is to have a \lq super-TAC\rq\ operating at a higher level than the canonical scientific category panels. A super-TAC would consider ideas from the perspective of a multi-observatory approach and specifically consider the merits of the proposed simultaneity of an observation. This is already possible to a certain extent with multi-observatory proposals allowed for some calls, albeit for relatively small amounts of time and without the possibility of ToOs. But a true multi-observatory campaign for responding to transient events remains hard to organise. A super-TAC could potentially operate on its own timeline, separate from the proposal deadlines of other observatories which are very often not synchronised with each other anyway. Such a super-TAC would also circumvent any question associated with which observatory to consider as the \lq primary\rq\ observatory for proposal submission. For true multi-wavelength science, there is often, by definition, no preference of one wavelength over another, and hence no primary observatory nor primary TAC. 

Key open questions to address in considering a super-TAC which are beyond the scope of this paper are:

\noindent {\it Organisation:} an obvious and important question is who would run such a TAC? One possibility is that there is nothing centralised or attached to a single organisation/institution but members are distributed around the globe. This distribution would also allow $\sim$24~hr coverage and therefore a fast response assuming that enough members of the TAC are available at any given time to ensure a level of impartiality. Such TAC members would of course have to be approved and supported by each observatory and any approved proposal based on science credentials would still require consultation with those observatories to check feasibility, particularly for fast-turnaround.

\noindent {\it Scale:} intrinsically connected to the above point is scale -- how many instruments should be involved and how much time on each observatory would be available to the TAC? It is likely that any such attempt would need to start with the major international instruments -- in particular those instruments whose waveband is difficult to cover: X-rays satellites, IR satellites, mid-IR ground-based observatories and mm facilities. Large ground-based facilities are also likely to be crucial components simply due to their reach to more distant objects (e.g. TDEs, GRBs, FRBs etc).

\noindent {\it Scientific threshold:} the expected deluge of new transients following the introduction of sensitive all-sky survey instruments (e.g. LSST, SKA - Figure 15) and subsequent demand placed on coordinated observing, coupled with the inherent difficulty faced in such coordination, requires that any science gain from such an endeavour be significant (and therefore offset the considerable effort). Whilst the potential science gains could be significant (see previous sections), having a number of reviewers on the super-TAC will naturally filter out weaker science cases.

Another possible solution to the issues of scheduling campaigns across present instruments (besides a super-TAC) would be to allow proposals to be submitted one or two cycles {\em in advance} of observation or programs that last for more than one cycle (for instance {\it XMM-Newton} ToOs remain valid if not triggered for 3 years). This would allow for other supporting observations to be proposed and approved in a timely manner, and would overcome the difficulty associated with unsynchronised proposal deadlines. 

Alternatively, more time could be allocated to larger joint facility proposals (such as the expanded joint proposal time made recently available via the {\it Chandra X-ray observatory} 
 - which can be submitted two cycles in advance to allow additional, coordinated observations to be obtained) although to study rapid, unexpected transients, such approved time would have to accommodate ToOs. Such an approach is likely to be valuable in the short term (for instance this would have made studying V404 Cygni far easier), however, MoUs do not exist between all instruments (even the major ones) and a larger coordinated strategy that all sizes of facilities can contribute to should be the final aim. Additionally - and inherently connected with joint scheduling - is the need for a pan-observatory visibility tool to allow PIs to determine which instruments can be used in a coordinated campaign. Such tools already exist for individual observatories (e.g. via ESA or HEASARC) and so combining the relevant information is a relatively straightforward task to implement should all observatories provide assistance.

{\it Slaved facilities.} A possible shortcut to avoid the need for a super-TAC altogether is to hard-link the scheduling of two or more facilities, as to make them act as a single multi-wavelength observatory. This is something that has already been implemented in some cases, one of the best-known examples being the Meer Licht optical telescope, which provides real-time optical coverage of the sky as observed by the radio telescope array MeerKAT. The small robotic optical telescope has the same instantaneous field-of-view as the radio array, and is fully dedicated to it, effectively acting as a `slaved' facility. Other similar setups are planned across the world, and it would be relatively easy to adopt this approach with multiple connected facilities, eventually covering an enormous fraction of the electromagnetic range. This approach has obvious advantages, as it simplifies most of the aspects mentioned above, especially for survey-dedicated facilities. The case for pointed observations is perhaps less clear-cut, a possible drawback being that not all targets (and/or science cases) need coverage at all wavelengths, which would eventually waste observing time at some of the involved facilities. A hybrid approach could perhaps solve this where observers could renounce some of the involved facilities, which would leave the door open to the use of these facilities by alternative programs.
This can be considered somewhat of an inverted philosophy with respect to the current one: whereas today different facilities can decide to dedicate a fraction of their observing time to multi-observatory programs, instead we would have a number of hard-linked, slaved facilities, each dedicating some fraction of their remaining observing time to simpler, single-observatory programs.

{\it Wider-band single facilities.} A somewhat extreme case of the above approach is represented by broad-band instruments. This is an old and well known approach, with many succesful examples (for example the currently active optical-to-IR instruments: X-Shooter on the VLT and GROND on the ESO 2.2m), and others are planned (for example the broad-band EVN``BRAND" receiver). The pros and cons of this philosophy are clear and similar to the ones mentioned above, with of course additional intrinsic limitations in bandwidth, as photons of very different energy need to be collected with very different technologies. An obvious solution is of course a multi-instrument satellite, which can offer a much broader energy coverage but prevents the different instruments being used separately by different simultaneous programs. Here further considerations apply, including the prosaic ones regarding costs, which go largely beyond the scope of this paper.

\item[--] \underline{Fast coordination.} For coordinating observations of `fast' (i.e. rapidly varying) transients (e.g. GRBs, XRBs, TDEs, GWs and FRBs) robotic networks are the most practical way forward. We are moving towards an era of not only automated detection of new transient events, but also automatic source classification with intelligent algorithmic agents. The human element enters at the initial stage of setting up the triggering and observation override criteria, but all observations (both primary and follow-up coordination) then occur without human intervention through virtual observatory (VO)Event communication between systems. These systems are easier to implement in small telescopes and the GRB community has a long history of making this work. However, larger facilities have also made bold steps in this direction, e.g. the Rapid Response Mode interruptions in-force at ESO and the VLT since 2003 (Vreeswijk et al. 2010) and at lower (radio) frequencies, the AMI (Anderson et al. 2014, Anderson et al., submitted), the Murchison Widefield Array (MWA) (which now triggers on GRBs: Kaplan et al. 2015) and the highly sensitive Australia Telescope Compact Array (ATCA). 

Another key component in improving efficient coordination for studies of fast transients (although it is also valuable for the study of objects which vary slowly) is the creation of a network of astronomers with access to time on a variety of instruments and expertise. This can be entirely informal and, as we discuss in the final section, we have made some progress towards this goal.  

\item[--] \underline{Real-time alerts.} Related to the requirement of fast, effective coordination is the proviso of real-time alerts from all-sky monitors which can then be used to trigger campaigns. As an example, GCN, GAIA, and ASASSN all send their real-time alerts to VOEvents. It is the VOEvents which then robotically trigger other instruments including AMI (and soon MWA). As the software is in place to robotically trigger on new transients, encouraging more facilities to send out VOEvent alerts is a sensible approach to ensure rapid follow-up of well known and newly discovered transients. 

\item[--] \underline{Access to expertise.} With regard to enabling observations by non-experts, one possible solution may be to have a few ready-made Phase 2 templates available, already filled out and tailored according to specific science motivations (e.g. pre-filled templates for fast timing, for bright source polarimetry, for faint object spectroscopy etc.). Observatory staff are very often better and more efficient at handling these tasks than the users. 

Support for interpretation of scientific data was also identified as severely lacking at present. The perception is that interpretation is not as novel or exciting as creation of new data. Interpretation efforts then suffer from lack of manpower, but are equally important. Interpretation, training in astro-statistics and development of requisite numerical techniques needs to be a part of the formative discussions of any campaign, rather than follow on at the end.

\item[--] \underline{Community action.} It is the obligation of the multi-wavelength variability community to effectively advertise the novel science that is enabled through simultaneous observations. Reaching out to astrophysics sub-communities with similar potential interests and operational difficulties could prove useful. Convincing non-experts of the importance of certain observations is usually much harder, but also much more important, than preaching to the converted.

\item[--] \underline{Involvement of theorists.}  There is a tendency for theorists and observers to work in a vacuum of their own creation with very little overlap. However, interpreting data relies on the contribution of theorists - as our quality of data and ability to co-ordinate across multiple bands improves, such scientific collaboration must become even more commonplace if we wish to make progress. Additionally -- and crucially -- to motivate the next generation of instruments, observers must consider what theory would predict: the best example of this in recent times is undoubtedly A-LIGO and the discovery of gravitational waves.

\end{itemize}

\section{Concluding remarks and SMARTNet}

The workshop in Leiden was a step in the right direction but only the first of many. Since the meeting took place, the {\it Chandra X-ray observatory} has expanded the time available via joint proposals, and now allows users to apply for increased amounts of time on several instruments (e.g. {\it HST}, {\it NuSTAR}, {\it Swift} and {\it XMM-Newton}) whilst still being able to apply for time on NRAO and NOAO facilities within the {\bf same} campaign. Whilst this does not guarantee simultaneity it is undoubtedly a sign that, at a time when `mission legacy' is becoming the new by-word, the community -- and those in positions of decision making power -- are becoming increasingly aware of the issues we have discussed in this white paper. 

Perhaps the most tangible outcome from the workshop was the creation of a network to facilitate coordination of simultaneous observations focused around individual campaigns, with minimal (ideally, zero) levels of bureaucracy and without the need for formal agreements amongst participants. After the meeting, the name \lq SMARTNet\rq\ (Simultaneous Multi-wavelength Astronomy Research in Transients Network) was agreed upon through a vote. SMARTNet\footnotemark\footnotetext{http://www.isdc.unige.ch/SmartNet/} is free for anyone to join and will provide a valuable resource of expertise in the use of multiple instruments across multiple observing bands as well as the employment of multiple techniques. We are hopeful that this network will develop to allow people to share data, instrument access, and thereby lay the foundations for improved observing campaigns of transients of all natures and speeds. 

\section{Acknowledgements}

We would like to thank the anonymous referee for their time and effort in providing comments which have helped improve the review. We also recognise and thank the Lorentz Center in Leiden for allowing us to hold the meeting and their staff for their tireless help in making the meeting such a success. We also thank Helen Russell for useful discussions. MJM and PG appreciate support from Ernest Rutherford STFC fellowships. PC acknowledges support by a Marie Curie FP7-Reintegration-Grant under contract no. 2012-322259. 

\section{References}

\begin{enumerate}
\item Abbott B.~P., et al., 2016, PhRvL, 116, 061102
\vspace{-0.3cm} 
\item Abbott B.~P., et al., 2016, ApJ, 826, L13
\vspace{-0.3cm} 
\item Abdo A.~A., et al., 2010, Natur, 463, 919
\vspace{-0.3cm}
\item Ackermann, M. et al. 2014, Sci 343, 42
\vspace{-0.3cm} 
\item Ackermann M., et al., 2016, ApJ, 824, L20 
\vspace{-0.3cm} 
\item Alexander T., 2012, EPJWC, 39, 05001
\vspace{-0.3cm} 
\item Anderson G.~E., et al., 2014, MNRAS, 440, 2059
\vspace{-0.3cm}
\item Antonucci R.~R.~J., Miller J.~S., 1985, ApJ, 297, 621
\vspace{-0.3cm} 
\item Antonucci R., 1993, ARA\&A, 31, 473
\vspace{-0.3cm} 
\item Arnett W.~D., 1980, ApJ, 237, 541
\vspace{-0.3cm} 
\item Arnett W.~D., 1982, ApJ, 253, 785
\vspace{-0.3cm} 

\item Bachetti M., et al., 2014, Natur, 514, 202
\vspace{-0.3cm} 
\item Bailey J., 1980, MNRAS, 190, 119 
\vspace{-0.3cm}
\item Bassa C.~G., et al., 2016, MNRAS, 463, L36
\vspace{-0.3cm} 
\item Becker C.~M., Remillard R.~A., Rappaport S.~A., McClintock J.~E., 1998, ApJ, 506, 880
\vspace{-0.3cm} 
\item Begelman M., Fabian A. \& Rees M. 2008, MNRAS 384, L19
\vspace{-0.3cm} 
\item Berger E., Kulkarni S.~R., Chevalier R.~A., 2002, ApJ, 577, L5 
\vspace{-0.3cm} 
\item Belloni T.~M., Motta S.~E., 2016, ASSL, 440, 61 
\vspace{-0.3cm} 
\item Bernardini F., Cackett E.~M., Brown E.~F., D'Angelo C., Degenaar N., Miller J.~M., Reynolds M., Wijnands R., 2013, MNRAS, 436, 2465
\vspace{-0.3cm}
\item Bernardini F., Russell D.~M., Kolojonen K.~I.~I., Stella L., Hynes R.~I., Corbel S., 2016, ApJ, 826, 149
\vspace{-0.3cm}
\item Bj{\"o}rnsson C.-I., Fransson C., 2004, ApJ, 605, 823
\vspace{-0.3cm} 
\item Blandford R.~D., Znajek R.~L., 1977, MNRAS, 179, 433 
\vspace{-0.3cm} 
\item Blandford R.~D., Begelman M.~C., 1999, MNRAS, 303, L1
\vspace{-0.3cm} 
\item Bloom J.~S., et al., 2011, Sci, 333, 203 
\vspace{-0.3cm}
\item Bogd{\'a}n {\'A}., Goulding A.~D., 2015, ApJ, 800, 124
\vspace{-0.3cm} 
\item Brightman M., et al., 2016, ApJ, 816, 60
\vspace{-0.3cm} 
\item Burrows D.~N., et al., 2011, Natur, 476, 421
\vspace{-0.3cm} 
\item Byckling K., Osborne J.~P., Wheatley P.~J., Wynn G.~A., Beardmore A., Braito V., Mukai K., West R.~G., 2009, MNRAS, 399, 1576
\vspace{-0.3cm} 

\item Cannizzo J.~K., Kenyon S.~J., 1987, ApJ, 320, 319 
\vspace{-0.3cm} 
\item Cao Y., et al., 2015, Natur, 521, 328
\vspace{-0.3cm} 
\item Camilo F., Ransom S.~M., Halpern J.~P., Reynolds J., Helfand D.~J., Zimmerman N., Sarkissian J., 2006, Natur, 442, 892
\vspace{-0.3cm} 
\item Casares J., Jonker P.~G., 2014, SSRv, 183, 223 
\vspace{-0.3cm} 
\item Casella P., et al., 2010, MNRAS, 404, L21 
\vspace{-0.3cm} 
\item Chakraborti S., et al., 2015, ApJ, 805, 187 
\vspace{-0.3cm} 
\item Chatterjee R., et al., 2009, ApJ, 704, 1689
\vspace{-0.3cm} 
\item Chatterjee R., et al., 2011, ApJ, 734, 43 
\vspace{-0.3cm} 
\item Chen X., Madau P., Sesana A., Liu F.~K., 2009, ApJ, 697, L149
\vspace{-0.3cm} 
\item Cheung C.~C., et al., 2016, ApJ, 826, 142 
\vspace{-0.3cm} 
\item Chevalier R.~A., 1982, ApJ, 259, 302
\vspace{-0.3cm} 
\item Chevalier R.~A., 1998, ApJ, 499, 810 
\vspace{-0.3cm}
\item Chevalier R.~A., Fransson C., 2006, ApJ, 651, 381
\vspace{-0.3cm} 
\item Chevalier R.~A., Fransson C., Nymark T.~K., 2006, ApJ, 641, 1029
\vspace{-0.3cm} 
\item Chomiuk L., et al., 2016, ApJ, 821, 119 
\vspace{-0.3cm} 
\item Chomiuk L., et al., 2012, ApJ, 750, 164
\vspace{-0.3cm} 
\item Colbert E.~J.~M., Mushotzky R.~F., 1999, ApJ, 519, 89
\vspace{-0.3cm} 
\item Colgate S.~A., McKee C., 1969, ApJ, 157, 623
\vspace{-0.3cm} 
\item Connors R.~M.~T., et al., 2017, MNRAS, 466, 4121 
\vspace{-0.3cm} 
\item Coppejans D.~L., et al., 2016, MNRAS, 
\vspace{-0.3cm} 
\item Coppejans D.~L., K{\"o}rding E.~G., Miller-Jones J.~C.~A., Rupen M.~P., Knigge C., Sivakoff G.~R., Groot P.~J., 2015, MNRAS, 451, 3801
\vspace{-0.3cm} 
\item Cseh D., et al., 2015, MNRAS, 446, 3268 
\vspace{-0.3cm} 
\item Cseh D., et al., 2015, MNRAS, 452, 24 
\vspace{-0.3cm} 
\item Cseh D., et al., 2014, MNRAS, 439, L1 
\vspace{-0.3cm} 
\item Curran P.~A., Chaty S., Zurita Heras J.~A., 2012, A\&A, 547, A41 
\vspace{-0.3cm} 

\item Davis S.~W., Narayan R., Zhu Y., Barret D., Farrell S.~A., Godet O., Servillat M., Webb N.~A., 2011, ApJ, 734, 111
\vspace{-0.3cm} 
\item Deller A.~T., et al., 2015, ApJ, 809, 13 
\vspace{-0.3cm} 
\item Dibi S., Markoff S., Belmont R., Malzac J., Barri{\`e}re N.~M., Tomsick J.~A., 2014, MNRAS, 441, 1005
\vspace{-0.3cm} 
\item Dibi S., Markoff S., Belmont R., Malzac J., Neilsen J., Witzel G., 2016, MNRAS, 461, 552
\vspace{-0.3cm} 
\item Dodds-Eden K., et al., 2011, ApJ, 728, 37
\vspace{-0.3cm} 
\item Done C., Gierli{\'n}ski M., Kubota A., 2007, A\&ARv, 15, 1 
\vspace{-0.3cm} 
\item Donnarumma I., Rossi E.~M., 2015, ApJ, 803, 36
\vspace{-0.3cm} 
\item Durant M., Gandhi P., Shahbaz T., Fabian A.~P., Miller J., Dhillon V.~S., Marsh T.~R., 2008, ApJ, 682, L45
\vspace{-0.3cm} 
\item Durant M., et al., 2011, MNRAS, 410, 2329 
\vspace{-0.3cm} 

\item Echevarr{\'{\i}}a J., Jones D., 1983, RMxAA, 5, 301 
\vspace{-0.3cm} 
\item Edelson R., et al., 2015, ApJ, 806, 129 
\vspace{-0.3cm} 
\item Eracleous M., Horne K., 1996, ApJ, 471, 427 
\vspace{-0.3cm} 
\item Esquej P., et al., 2008, A\&A, 489, 543 
\vspace{-0.3cm} 
\item Evans C.~R., Kochanek C.~S., 1989, ApJ, 346, L13 
\vspace{-0.3cm} 

\item Fabian A.~C., 2009, arXiv, arXiv:0908.2784 
\vspace{-0.3cm}
\item Fabian A.~C., 2012, ARA\&A, 50, 455
\vspace{-0.3cm} 
\item Fabrika S., 2004, ASPRv, 12, 1
\vspace{-0.3cm}
\item Falcke H., Goss W.~M., Matsuo H., Teuben P., Zhao J.-H., Zylka R., 1998, ApJ, 499, 731  
\vspace{-0.3cm}
\item Falcke H., K{\"o}rding E., Markoff S., 2004, A\&A, 414, 895
\vspace{-0.3cm}
\item Farrell S.~A., Webb N.~A., Barret D., Godet O., Rodrigues J.~M., 2009, Natur, 460, 73
\vspace{-0.3cm}
\item Farrell S.~A., et al., 2012, ApJ, 747, L13 
\vspace{-0.3cm}
\item Fausnaugh M.~M., et al., 2016, ApJ, 821, 56 
\vspace{-0.3cm}
\item Fender R.~P., Homan J., Belloni T.~M., 2009, MNRAS, 396, 1370
\vspace{-0.3cm}
\item Fender R.~P., Belloni T.~M., Gallo E., 2004, MNRAS, 355, 1105
\vspace{-0.3cm} 
\item Fender R.~P., Gallo E., Jonker P.~G., 2003, MNRAS, 343, L99 
\vspace{-0.3cm} 
\item Fender R.~P., 2003, MNRAS, 340, 1353 
\vspace{-0.3cm}
\item Fender R.~P., Kuulkers E., 2001, MNRAS, 324, 923
\vspace{-0.3cm} 
\item Feng H., Soria R., 2011, NewAR, 55, 166 
\vspace{-0.3cm} 
\item Ferrarese L., Merritt D., 2000, ApJ, 539, L9 
\vspace{-0.3cm}
\item Filippenko A.~V., 1997, ARA\&A, 35, 309
\vspace{-0.3cm} 
\item Fuerst F., et al., 2016, ApJ, 831, L14
\vspace{-0.3cm}

\item Galama T.~J., et al., 1998, Natur, 395, 670 
\vspace{-0.3cm} 
\item Gallo E., Migliari S., Markoff S., Tomsick J.~A., Bailyn C.~D., Berta S., Fender R., Miller-Jones J.~C.~A., 2007, ApJ, 670, 600 
\vspace{-0.3cm} 
\item Gandhi P., et al., 2008, MNRAS, 390, L29 
\vspace{-0.3cm} 
\item Gandhi P., 2009, ApJ, 697, L167
\vspace{-0.3cm} 
\item Gandhi P., et al., 2010, MNRAS, 407, 2166 
\vspace{-0.3cm} 
\item Gandhi P., et al., 2011, ApJ, 740, L13 
\vspace{-0.3cm} 
\item Gandhi P., et al., 2016, MNRAS, 459, 554
\vspace{-0.3cm}
\item Gebhardt K., et al., 2000, ApJ, 539, L13 
\vspace{-0.3cm}
\item Gehrels N., Ramirez-Ruiz E., Fox D.~B., 2009, ARA\&A, 47, 567
\vspace{-0.3cm}
\item Genzel R., Eisenhauer F., Gillessen S., 2010, RvMP, 82, 3121
\vspace{-0.3cm} 
\item Gezari S., et al., 2008, ApJ, 676, 944-969 
\vspace{-0.3cm} 
\item Gezari S., et al., 2010, ApJ, 720, L77 
\vspace{-0.3cm} 
\item Gierli{\'n}ski M., Done C., Page K., 2009, MNRAS, 392, 1106 
\vspace{-0.3cm} 
\item Gierli{\'n}ski M., Zdziarski A.~A., Poutanen J., Coppi P.~S., Ebisawa K., Johnson W.~N., 1999, MNRAS, 309, 496
\vspace{-0.3cm}
\item Gladstone J.~C., Roberts T.~P., Done C., 2009, MNRAS, 397, 1836 
\vspace{-0.3cm} 
\item Graham, M.~L., Nugent, P.~E., Sullivan, M., et al. 2015, MNRAS, 454, 1948 
\vspace{-0.3cm}
\item Granot, J. \& van der Horst, A. J. 2014, PASA 31, 8 
\vspace{-0.3cm} 
\item Green D.~A., Stephenson F.~R., 2003, LNP, 598, 7 
\vspace{-0.3cm}

\item Hackwell J.~A., Grasdalen G.~L., Gehrz R.~D., Cominsky L., Lewin W.~H.~G., van Paradijs J., 1979, ApJ, 233, L115
\vspace{-0.3cm} 
\item Hameury J.-M., Lasota J.-P., Knigge C., K{\"o}rding E.~G., 2017, A\&A, 600, A95
\vspace{-0.3cm} 
\item Hancock P.~J., Gaensler B.~M., Murphy T., 2011, ApJ, 735, L35 
\vspace{-0.3cm}
\item Hayashida M., et al., 2015, ApJ, 807, 79
\vspace{-0.3cm}
\item Hill A.~B., et al., 2011, MNRAS, 415, 235
\vspace{-0.3cm} 
\item Hjorth, J. et al. 2012, ApJ, 756, 187 
\vspace{-0.3cm} 
\item Ho L.~C., 1999, ApJ, 516, 672 
\vspace{-0.3cm} 
\item Homan J., Neilsen J., Allen J.~L., Chakrabarty D., Fender R., Fridriksson J.~K., Remillard R.~A., Schulz N., 2016, ApJ, 830, L5 
\vspace{-0.3cm} 
\item Hynes R.~I., et al., 2003, MNRAS, 345, 292 
\vspace{-0.3cm} 
\item Hynes R.~I., Horne K., O'Brien K., Haswell C.~A., Robinson E.~L., King A.~R., Charles P.~A., Pearson K.~J., 2006, ApJ, 648, 1156
\vspace{-0.3cm}
\item Hynes R.~I., Bradley C.~K., Rupen M., Gallo E., Fender R.~P., Casares J., Zurita C., 2009, MNRAS, 399, 2239 
\vspace{-0.3cm} 

\item Ichimaru S., 1977, ApJ, 214, 840
\vspace{-0.3cm} 
\item Ingram A., van der Klis M., Middleton M., Altamirano D., Uttley P., 2017, MNRAS, 464, 2979
\vspace{-0.3cm}
\item Ingram A., van der Klis M., Middleton M., Done C., Altamirano D., Heil L., Uttley P., Axelsson M., 2016, MNRAS, 461, 1967
\vspace{-0.3cm} 
\item Israel G.~L., et al., 2017, MNRAS, 466, L48
\vspace{-0.3cm} 
\item Israel G.~L., et al., 2017, Science, 355, 817
\vspace{-0.3cm} 
\item Iwamoto K., et al., 1998, Natur, 395, 672
\vspace{-0.3cm} 

\item Japelj, J. et al. 2014, ApJ 785, 84
\vspace{-0.3cm} 
\item Jiang Y.-F., Stone J.~M., Davis S.~W., 2014, ApJ, 796, 106 
\vspace{-0.3cm} 
\item Jin C., Ward M., Done C., Gelbord J., 2012, MNRAS, 420, 1825 
\vspace{-0.3cm} 
\item Jorstad S.~G., et al., 2013, ApJ, 773, 147
\vspace{-0.3cm}

\item Kaaret P., Feng H., Wong D.~S., Tao L., 2010, ApJ, 714, L167 
\vspace{-0.3cm} 
\item Kaaret P., Feng H., Roberts T.~P., 2017, arXiv, arXiv:1703.10728
\vspace{-0.3cm} 
\item Kaastra J.~S., et al., 2014, Sci, 345, 64
\vspace{-0.3cm}
\item Kalamkar M., Casella P., Uttley P., O'Brien K., Russell D., Maccarone T., van der Klis M., Vincentelli F., 2016, MNRAS, 460, 3284 
\vspace{-0.3cm} 
\item Kaplan D.~L., et al., 2015, ApJ, 814, L25
\vspace{-0.3cm} 
\item Kanbach G., Straubmeier C., Spruit H.~C., Belloni T., 2001, Natur, 414, 180
\vspace{-0.3cm} 
\item Kara E., Miller J.~M., Reynolds C., Dai L., 2016, Natur, 535, 388
\vspace{-0.3cm}
\item Kashiyama K., Murase K., Horiuchi S., Gao S., M{\'e}sz{\'a}ros P., 2013, ApJ, 769, L6 
\vspace{-0.3cm}
\item Keane E.~F., et al., 2016, Natur, 530, 453 
\vspace{-0.3cm} 
\item Keek L., Wolf Z., Ballantyne D.~R., 2016, ApJ, 826, 79 
\vspace{-0.3cm} 
\item Khabibullin I., Sazonov S., 2014, MNRAS, 444, 1041
\vspace{-0.3cm} 
\item Kimura M., et al., 2016, Natur, 529, 54
\vspace{-0.3cm} 
\item King A.~R., 2009, MNRAS, 393, L41 
\vspace{-0.3cm}
\item King A.~L., Miller J.~M., Raymond J., Reynolds M.~T., Morningstar W., 2015, ApJ, 813, L37 
\vspace{-0.3cm} 
\item King A., Lasota J.-P., Klu{\'z}niak W., 2017, MNRAS, 468, L59
\vspace{-0.3cm} 
\item Kinney A.~L., Schmitt H.~R., Clarke C.~J., Pringle J.~E., Ulvestad J.~S., Antonucci R.~R.~J., 2000, ApJ, 537, 152 
\vspace{-0.3cm} 
\item Klu{\'z}niak W., Lasota J.-P., 2015, MNRAS, 448, L43 
\vspace{-0.3cm} 
\item K{\"o}rding E.~G., Jester S., Fender R., 2006, MNRAS, 372, 1366
\vspace{-0.3cm}
\item K{\"o}rding E., Rupen M., Knigge C., Fender R., Dhawan V., Templeton M., Muxlow T., 2008, Sci, 320, 1318 
\vspace{-0.3cm} 
\item Knigge C., Sivakoff G.~R., Kuulkers E., Altamirano D., Neilsen J., 2015, ATel, 7773
\vspace{-0.3cm} 
\item Knigge C., Marsh T.~R., Sivakoff G.~R., Altamirano D., Hern{\'a}ndez Santisteban J.~V., Shaw A., Charles P.~A., Gandhi P., 2015, ATel, 7735
\vspace{-0.3cm} 
\item Komossa S., 2015, JHEAp, 7, 148 
\vspace{-0.3cm} 
\item Komossa S., Greiner J., 1999, A\&A, 349, L45 
\vspace{-0.3cm} 
\item Koss M.~J., et al., 2016, ApJ, 825, 85 
\vspace{-0.3cm} 
\item Kouveliotou, C. et al. 1993, ApJ 413L, 101
\vspace{-0.3cm} 
\item Kulkarni S.~R., et al., 1998, Natur, 395, 663
\vspace{-0.3cm} 
\item Kuulkers E., 2015, ATel, 7695
\vspace{-0.3cm} 

\item Laskar T., et al., 2013, ApJ, 776, 119
\vspace{-0.3cm} 
\item Lasota J.-P., 2001, NewAR, 45, 449 
\vspace{-0.3cm} 
\item Lasota J.-P., 2016, ASSL, 440, 1 
\vspace{-0.3cm} 
\item Lewin W.~H.~G., van Paradijs J., Taam R.~E., 1995, xrbi.nasa, 175 
\vspace{-0.3cm} 
\item Livio M., 1999, PhR, 311, 225
\vspace{-0.3cm} 
\item Lohfink A.~M., et al., 2013, ApJ, 772, 83
\vspace{-0.3cm} 
\item Lorimer D.~R., Bailes M., McLaughlin M.~A., Narkevic D.~J., Crawford F., 2007, Sci, 318, 777
\vspace{-0.3cm} 
\item Luo B., Brandt W.~N., Steffen A.~T., Bauer F.~E., 2008, ApJ, 674, 122-132
\vspace{-0.3cm} 

\item Maccarone T.~J., 2003, A\&A, 409, 697 
\vspace{-0.3cm}
\item Maccarone T.~J., Coppi P.~S., 2003, A\&A, 399, 1151 
\vspace{-0.3cm}
\item Maccarone T.~J., Coppi P.~S., 2003, MNRAS, 338, 189 
\vspace{-0.3cm}
\item Macquart J.~P., et al., 2015, aska.conf, 55
\vspace{-0.3cm}
\item Maeda K., 2013, ApJ, 762, 14 
\vspace{-0.3cm}
\item Magorrian J., et al., 1998, AJ, 115, 2285
\vspace{-0.3cm}
\item Maitra D., Bailyn C.~D., 2008, ApJ, 688, 537-549 
\vspace{-0.3cm} 
\item Maksym W.~P., Ulmer M.~P., Eracleous M., 2010, ApJ, 722, 1035
\vspace{-0.3cm} 
\item Margutti R., Parrent J., Kamble A., Soderberg A.~M., Foley R.~J., Milisavljevic D., Drout M.~R., Kirshner R., 2014, ApJ, 790, 52
\vspace{-0.3cm} 
\item Margutti R., et al., 2017, ApJ, 835, 140
\vspace{-0.3cm}
\item Margutti R., et al., 2012, ApJ, 751, 134 
\vspace{-0.3cm} 
\item Marion G.~H., et al., 2016, ApJ, 820, 92
\vspace{-0.3cm} 
\item Markoff S., et al., 2015, ApJ, 812, L25 
\vspace{-0.3cm} 
\item Marscher A.~P., Jorstad S.~G., G{\'o}mez J.-L., Aller M.~F., Ter{\"a}sranta H., Lister M.~L., Stirling A.~M., 2002, Natur, 417, 625
\vspace{-0.3cm} 
\item Marscher A., Jorstad S.~G., Larionov V.~M., Aller M.~F., L{\"a}hteenm{\"a}ki A., 2011, JApA, 32, 233
\vspace{-0.3cm}
\item Matsuoka M., et al., 1984, ApJ, 283, 774
\vspace{-0.3cm}
\item Mattei J.~A., Mauche C., Wheatley P.~J., 2000, JAVSO, 28, 160
\vspace{-0.3cm}
\item McClintock J.~E., Narayan R., Steiner J.~F., 2014, SSRv, 183, 295
\vspace{-0.3cm} 
\item McClintock J.~E., Shafee R., Narayan R., Remillard R.~A., Davis S.~W., Li L.-X., 2006, ApJ, 652, 518
\vspace{-0.3cm} 
\item McConnell M.~L., et al., 2000, ApJ, 543, 928 
\vspace{-0.3cm} 
\item McGowan K.~E., Priedhorsky W.~C., Trudolyubov S.~P., 2004, ApJ, 601, 1100 
\vspace{-0.3cm} 
\item McHardy I.~M., et al., 2014, MNRAS, 444, 1469 
\vspace{-0.3cm} 
\item McNamara B.~R., Kazemzadeh F., Rafferty D.~A., B{\^i}rzan L., Nulsen P.~E.~J., Kirkpatrick C.~C., Wise M.~W., 2009, ApJ, 698, 594 
\vspace{-0.3cm} 
\item Mereghetti S., 2008, A\&ARv, 15, 225 
\vspace{-0.3cm} 
\item Middleton M.~J., King A., 2017, MNRAS, 470, L69 
\vspace{-0.3cm} 
\item Middleton M.~J., Parker M.~L., Reynolds C.~S., Fabian A.~C., Lohfink A.~M., 2016, MNRAS, 457, 1568 
\vspace{-0.3cm} 
\item Middleton M., Done C., Gierli{\'n}ski M., 2007, MNRAS, 381, 1426
\vspace{-0.3cm} 
\item Middleton M., 2016, ASSL, 440, 99
\vspace{-0.3cm} 
\item Middleton M.~J., et al., 2013, Natur, 493, 187
\vspace{-0.3cm} 
\item Middleton M.~J., Miller-Jones J.~C.~A., Fender R.~P., 2014, MNRAS, 439, 1740 
\vspace{-0.3cm} 
\item Middleton M.~J., Walton D.~J., Roberts T.~P., Heil L., 2014, MNRAS, 438, L51
\vspace{-0.3cm} 
\item Middleton M.~J., Walton D.~J., Fabian A., Roberts T.~P., Heil L., Pinto C., Anderson G., Sutton A., 2015, MNRAS, 454, 3134
\vspace{-0.3cm} 
\item Middleton M.~J., Ingram A.~R., 2015, MNRAS, 446, 1312 
\vspace{-0.3cm} 
\item Migliari S., Fender R.~P., 2006, MNRAS, 366, 79 
\vspace{-0.3cm} 
\item Migliari S., Fender R.~P., Rupen M., Jonker P.~G., Klein-Wolt M., Hjellming R.~M., van der Klis M., 2003, MNRAS, 342, L67
\vspace{-0.3cm} 
\item Migliari S., Miller-Jones J.~C.~A., Russell D.~M., 2011, MNRAS, 415, 2407
\vspace{-0.3cm} 
\item Migliari S., Fender R.~P., Rupen M., Wachter S., Jonker P.~G., Homan J., van der Klis M., 2004, MNRAS, 351, 186
\vspace{-0.3cm} 
\item Miller J.~M., Raymond J., Fabian A., Steeghs D., Homan J., Reynolds C., van der Klis M., Wijnands R., 2006, Natur, 441, 953 
\vspace{-0.3cm} 
\item Miller-Jones J.~C.~A., Fender R.~P., Nakar E., 2006, MNRAS, 367, 1432
\vspace{-0.3cm} 
\item Miller-Jones J.~C.~A., Jonker P.~G., Dhawan V., Brisken W., Rupen M.~P., Nelemans G., Gallo E., 2009, ApJ, 706, L230 
\vspace{-0.3cm} 
\item Miller-Jones J.~C.~A., et al., 2010, ApJ, 716, L109
\vspace{-0.3cm} 
\item Mirabel I.~F., Rodr{\'{\i}}guez L.~F., 1994, Natur, 371, 46
\vspace{-0.3cm} 
\item Modjaz M., Liu Y.~Q., Bianco F.~B., Graur O., 2016, ApJ, 832, 108
\vspace{-0.3cm} 
\item Motch C., 1998, A\&A, 338, L13 
\vspace{-0.3cm} 
\item Motch C., Pakull M.~W., Soria R., Gris{\'e} F., Pietrzy{\'n}ski G., 2014, Natur, 514, 198
\vspace{-0.3cm}
\item Mushtukov A.~A., Suleimanov V.~F., Tsygankov S.~S., Poutanen J., 2015, MNRAS, 454, 2539
\vspace{-0.3cm} 

\item Narayan R., Piran T., 2012, MNRAS, 420, 604
\vspace{-0.3cm}
\item Narayan R., Yi I., 1994, ApJ, 428, L13 
\vspace{-0.3cm} 
\item Narayan R., Yi I., 1995, ApJ, 444, 231
\vspace{-0.3cm} 
\item Neilsen J., et al., 2015, ApJ, 799, 199
\vspace{-0.3cm} 
\item Neilsen J., Lee J.~C., 2009, Natur, 458, 481 
\vspace{-0.3cm} 

\item Ofek E.~O., et al., 2010, ApJ, 724, 1396
\vspace{-0.3cm}
\item Olling R.~P., et al., 2015, Natur, 521, 332 
\vspace{-0.3cm}
\item de O{\~n}a Wilhelmi E., et al., 2016, MNRAS, 456, 2647
\vspace{-0.3cm} 
\item Oruru B., Meintjes P.~J., 2014, EPJWC, 64, 07003 
\vspace{-0.3cm}
\item {\"O}zel F., Psaltis D., Narayan R., McClintock J.~E., 2010, ApJ, 725, 1918 
\vspace{-0.3cm} 

\item Pacciani L., Tavecchio F., Donnarumma I., Stamerra A., Carrasco L., Recillas E., Porras A., Uemura M., 2014, ApJ, 790, 45
\vspace{-0.3cm}
\item Pahari M., Gandhi P., Charles P.~A., Kotze M.~M., Altamirano D., Misra R., 2017, MNRAS, 469, 193
\vspace{-0.3cm}
\item Pakull M.~W., Mirioni L., 2003, RMxAC, 15, 197 
\vspace{-0.3cm} 
\item Papitto A., Torres D.~F., 2015, ApJ, 807, 33
\vspace{-0.3cm} 
\item Papitto A., Torres D.~F., Li J., 2014, MNRAS, 438, 2105 
\vspace{-0.3cm} 
\item Parker M.~L., et al., 2017, Natur, 543, 83 
\vspace{-0.3cm} 
\item Patat F., et al., 2007, Sci, 317, 924 
\vspace{-0.3cm} 
\item Patruno A., Maitra D., Curran P.~A., D'Angelo C., Fridriksson J.~K., Russell D.~M., Middleton M., Wijnands R., 2016, ApJ, 817, 100
\vspace{-0.3cm} 
\item Patruno A., et al., 2014, ApJ, 781, L3 
\vspace{-0.3cm} 
\item Pennucci T.~T., et al., 2015, ApJ, 808, 81 
\vspace{-0.3cm} 
\item Perlmutter S., et al., 1999, ApJ, 517, 565 
\vspace{-0.3cm} 
\item P{\'e}rez-Torres M.~A., et al., 2014, ApJ, 792, 38
\vspace{-0.3cm} 
\item Peterson B.~M., 2014, SSRv, 183, 253
\vspace{-0.3cm} 
\item Peterson B.~M., et al., 2002, ApJ, 581, 197
\vspace{-0.3cm} 
\item Petroff E., et al., 2015, MNRAS, 447, 246 
\vspace{-0.3cm}
\item Plotkin R.~M., Markoff S., Kelly B.~C., K{\"o}rding E., Anderson S.~F., 2012, MNRAS, 419, 267 
\vspace{-0.3cm} 
\item Pinto C., Middleton M.~J., Fabian A.~C., 2016, Natur, 533, 64
\vspace{-0.3cm} 
\item Piran, T. 2004, RvMP, 76, 1143
\vspace{-0.3cm} 
\item Piro L., et al., 1998, A\&A, 329, 906 
\vspace{-0.3cm} 
\item Ponti G., Fender R.~P., Begelman M.~C., Dunn R.~J.~H., Neilsen J., Coriat M., 2012, MNRAS, 422, 11
\vspace{-0.3cm} 
\item Poutanen J., Lipunova G., Fabrika S., Butkevich A.~G., Abolmasov P., 2007, MNRAS, 377, 1187
\vspace{-0.3cm} 
\item Poutanen J., Stern B., 2010, ApJ, 717, L118
\vspace{-0.3cm}

\item Rea N., Esposito P., 2011, ASSP, 21, 247 
\vspace{-0.3cm} 
\item Rees M.~J., 1988, Natur, 333, 523 
\vspace{-0.3cm} 
\item Rees M.~J., 1990, Sci, 247, 817 
\vspace{-0.3cm} 
\item Reid M.~J., McClintock J.~E., Steiner J.~F., Steeghs D., Remillard R.~A., Dhawan V., Narayan R., 2014, ApJ, 796, 2 
\vspace{-0.3cm} 
\item Reynolds C.~S., Young A.~J., Begelman M.~C., Fabian A.~C., 1999, ApJ, 514, 164
\vspace{-0.3cm} 
\item Riess A.~G., et al., 1998, AJ, 116, 1009
\vspace{-0.3cm} 
\item Ridgway S.~T., Matheson T., Mighell K.~J., Olsen K.~A., Howell S.~B., 2014, ApJ, 796, 53 
\vspace{-0.3cm}
\item Risaliti G., Nardini E., Elvis M., Brenneman L., Salvati M., 2011, MNRAS, 417, 178
\vspace{-0.3cm} 
\item Roberts T.~P., 2007, Ap\&SS, 311, 203 
\vspace{-0.3cm} 
\item Rodriguez J., et al., 2015, A\&A, 581, L9 
\vspace{-0.3cm} 
\item Rupen M.~P., Mioduszewski A.~J., Sokoloski J.~L., 2008, ApJ, 688, 559-567
\vspace{-0.3cm} 
\item Russell T.~D., Soria R., Miller-Jones J.~C.~A., Curran P.~A., Markoff S., Russell D.~M., Sivakoff G.~R., 2014, MNRAS, 439, 1390 
\vspace{-0.3cm} 
\item Russell D.~M., Gallo E., Fender R.~P., 2013, MNRAS, 431, 405 
\vspace{-0.3cm} 
\item Russell D.~M., et al., 2013, MNRAS, 429, 815 
\vspace{-0.3cm} 
\item Russell D.~M., Fender R.~P., Hynes R.~I., Brocksopp C., Homan J., Jonker P.~G., Buxton M.~M., 2006, MNRAS, 371, 1334
\vspace{-0.3cm} 
\item Rykoff E.~S., Cackett E.~M., Miller J.~M., 2010, ApJ, 719, 1993 
\vspace{-0.3cm} 

\item S{\c a}dowski A., Narayan R., McKinney J.~C., Tchekhovskoy A., 2014, MNRAS, 439, 503 
\vspace{-0.3cm} 
\item Sari, R., Piran, T. \& Narayan, R. 1998, ApJ 497, L17
\vspace{-0.3cm} 
\item Schreiber M.~R., Hameury J.-M., Lasota J.-P., 2003, A\&A, 410, 239
\vspace{-0.3cm} 
\item Schawinski K., et al., 2008, Sci, 321, 223 
\vspace{-0.3cm} 
\item Servillat M., Farrell S.~A., Lin D., Godet O., Barret D., Webb N.~A., 2011, ApJ, 743, 6
\vspace{-0.3cm} 
\item Shahbaz T., Russell D.~M., Zurita C., Casares J., Corral-Santana J.~M., Dhillon V.~S., Marsh T.~R., 2013, MNRAS, 434, 2696
\vspace{-0.3cm} 
\item Shakura N.~I., Sunyaev R.~A., 1973, A\&A, 24, 337
\vspace{-0.3cm} 
\item Skopal A., Tomov N.~A., Tomova M.~T., 2013, A\&A, 551, L10
\vspace{-0.3cm} 
\item Smith D.~M., Heindl W.~A., Markwardt C.~B., Swank J.~H., 2001, ApJ, 554, L41 
\vspace{-0.3cm}
\item Smith N., 2014, ARA\&A, 52, 487 
\vspace{-0.3cm} 
\item Smith N., Mauerhan J.~C., Prieto J.~L., 2014, MNRAS, 438, 1191
\vspace{-0.3cm} 
\item Soderberg A.~M., et al., 2008, Natur, 453, 469 
\vspace{-0.3cm} 
\item Soderberg A.~M., et al., 2010, Natur, 463, 513
\vspace{-0.3cm} 
\item Soderberg A.~M., Nakar E., Berger E., Kulkarni S.~R., 2006, ApJ, 638, 930
\vspace{-0.3cm} 
\item Sokoloski J.~L., Rupen M.~P., Mioduszewski A.~J., 2008, ApJ, 685, L137
\vspace{-0.3cm} 
\item Soria R., Kuntz K.~D., Winkler P.~F., Blair W.~P., Long K.~S., Plucinsky P.~P., Whitmore B.~C., 2012, ApJ, 750, 152
\vspace{-0.3cm} 
\item Spitler L.~G., et al., 2016, Natur, 531, 202
\vspace{-0.3cm}
\item Stanek K.~Z., et al., 2003, ApJ, 591, L17 
\vspace{-0.3cm} 
\item Stappers B.~W., et al., 2014, ApJ, 790, 39 
\vspace{-0.3cm} 
\item Steiner J.~F., McClintock J.~E., Narayan R., 2013, ApJ, 762, 104 
\vspace{-0.3cm} 
\item Steiner J.~F., Walton D.~J., Garc{\'{\i}}a J.~A., McClintock J.~E., Laycock S.~G.~T., Middleton M.~J., Barnard R., Madsen K.~K., 2016, ApJ, 817, 154 
\vspace{-0.3cm} 
\item Steiner J.~F., McClintock J.~E., Orosz J.~A., Remillard R.~A., Bailyn C.~D., Kolehmainen M., Straub O., 2014, ApJ, 793, L29 
\vspace{-0.3cm} 
\item Sternberg, A., Gal-Yam, A., Simon, J.~D., et al. 2011, Science, 333, 856 
\vspace{-0.3cm} 
\item Stobbart A.-M., Roberts T.~P., Wilms J., 2006, MNRAS, 368, 397
\vspace{-0.3cm} 
\item Strohmayer T., Bildsten L., 2006, csxs.book, 39, 113
\vspace{-0.3cm} 

\item Tam C.~R., Kaspi V.~M., van Kerkwijk M.~H., Durant M., 2004, ApJ, 617, L53
\vspace{-0.3cm} 
\item Tetarenko A.~J., et al., 2016, MNRAS, 460, 345
\vspace{-0.3cm} 
\item Toma, K., Ioka, K. \& Nakamura, T. 2008, ApJ 673, L123
\vspace{-0.3cm} 
\item Tombesi F., Mel{\'e}ndez M., Veilleux S., Reeves J.~N., Gonz{\'a}lez-Alfonso E., Reynolds C.~S., 2015, Natur, 519, 436
\vspace{-0.3cm} 
\item Truran J.~W., Arnett W.~D., Cameron A.~G.~W., 1967, CaJPh, 45, 2315
\vspace{-0.3cm} 
\item Turolla R., Zane S., Watts A.~L., 2015, RPPh, 78, 116901 
\vspace{-0.3cm}

\item Uttley P., Cackett E.~M., Fabian A.~C., Kara E., Wilkins D.~R., 2014, A\&ARv, 22, 72
\vspace{-0.3cm} 
\item Uttley P., Casella P., 2014, SSRv, 183, 453
\vspace{-0.3cm} 

\item Vadawale S.~V., Rao A.~R., Naik S., Yadav J.~S., Ishwara-Chandra C.~H., Pramesh Rao A., Pooley G.~G., 2003, ApJ, 597, 1023 
\vspace{-0.3cm} 
\item van Dyk S.~D., Weiler K.~W., Sramek R.~A., Rupen M.~P., Panagia N., 1994, ApJ, 432, L115 
\vspace{-0.3cm}
\item van den Heuvel E.~P.~J., 1992, eocm.rept,
\vspace{-0.3cm} 
\item van Paradijs J., 1981, A\&A, 103, 140 
\vspace{-0.3cm} 
\item van Paradijs J., McClintock J.~E., 1994, A\&A, 290, 133 
\vspace{-0.3cm} 
\item van Velzen S., Farrar G.~R., 2014, ApJ, 792, 53
\vspace{-0.3cm} 
\item Veledina A., Poutanen J., Vurm I., 2011, ApJ, 737, L17
\vspace{-0.3cm} 
\item Veledina A., Revnivtsev M.~G., Durant M., Gandhi P., Poutanen J., 2015, MNRAS, 454, 2855
\vspace{-0.3cm} 
\item Vreeswijk P.~M., Kaufer A., Spyromilio J., Schmutzer R., Ledoux C., Smette A., De Cia A., 2010, SPIE, 7737, 77370M
\vspace{-0.3cm} 

\item Walton D.~J., et al., 2016, ApJ, 826, L26 
\vspace{-0.3cm} 
\item Walton D.~J., et al., 2017, arXiv, arXiv:1705.10297
\vspace{-0.3cm} 
\item Wang X.-Y., Li Z., Waxman E., M{\'e}sz{\'a}ros P., 2007, ApJ, 664, 1026
\vspace{-0.3cm} 
\item Wang B., Han Z., 2012, NewAR, 56, 122
\vspace{-0.3cm} 
\item Wang Q.~D., et al., 2013, Sci, 341, 981
\vspace{-0.3cm} 
\item Webb N., et al., 2012, Sci, 337, 554 
\vspace{-0.3cm} 
\item Weiler K.~W., Panagia N., Montes M.~J., Sramek R.~A., 2002, ARA\&A, 40, 387 
\vspace{-0.3cm} 
\item Weiler K.~W., Williams C.~L., Panagia N., Stockdale C.~J., Kelley M.~T., Sramek R.~A., Van Dyk S.~D., Marcaide J.~M., 2007, ApJ, 671, 1959 
\vspace{-0.3cm} 
\item Wheatley P.~J., Mauche C.~W., Mattei J.~A., 2003, MNRAS, 345, 49 
\vspace{-0.3cm} 
\item Wiersema, K. 2014, Natur 509, 201
\vspace{-0.3cm} 
\item Wijers R. A. M. J. \& Galama, T. J. 1999, ApJ 523, 177
\vspace{-0.3cm} 
\item Wilkinson T., Uttley P., 2009, MNRAS, 397, 666 
\vspace{-0.3cm} 
\item Williams P.~K.~G., Berger E., 2016, ApJ, 821, L22
\vspace{-0.3cm}
\item Witzel G., et al., 2012, ApJS, 203, 18 
\vspace{-0.3cm} 
\item Woosley S.~E., Bloom J.~S., 2006, ARA\&A, 44, 507
\vspace{-0.3cm} 
\item Woosley, S. E. \& Bloom, J. S. 2007, ARA\&A 44, 507
\vspace{-0.3cm} 
\item Wu K., Soria R., Hunstead R.~W., Johnston H.~M., 2001, MNRAS, 320, 177
\vspace{-0.3cm}
\item Wynn G.~A., King A.~R., 1995, MNRAS, 275, 9 
\vspace{-0.3cm} 

\item Yonetoku, D. et al. 2012, ApJ 758, L1
\vspace{-0.3cm} 
\item Yuan F., Narayan R., 2014, ARA\&A, 52, 529 
\vspace{-0.3cm} 

\item Zauderer B.~A., et al., 2011, Natur, 476, 425
\vspace{-0.3cm} 
\item Zhu L., Di Stefano R., Wyrzykowski L., 2012, ApJ, 761, 118 
\vspace{-0.3cm} 
\end{enumerate}

\end{document}